\documentclass[useAMS,usenatbib]{mn2e}
\usepackage{epsf,graphicx, rotating}

\title[BL Lacs and radio galaxies]{On the relationship between BL
Lacertae objects and radio galaxies} 

\author[H. Landt \& H. E. Bignall]{
Hermine Landt$^1$\thanks{E-mail: hlandt@cfa.harvard.edu} and 
Hayley E. Bignall$^2$\thanks{Current address: Curtin University of Technology, 
GPO Box U1987, Perth, WA 6845, Australia} \\ 
$^1$Harvard-Smithsonian Center for Astrophysics,
60 Garden Street, Cambridge, MA 02138, USA \\ 
$^2$Joint Institute for VLBI in Europe, Postbus 2, NL-7990 AA
Dwingeloo, The Netherlands}

\begin{document}

\def\la{\mathrel{\hbox{\rlap{\hbox{\lower4pt\hbox{$\sim$}}}\hbox{$<$}}}}
\def\ga{\mathrel{\hbox{\rlap{\hbox{\lower4pt\hbox{$\sim$}}}\hbox{$>$}}}}

\font\sevenrm=cmr7
\def\Ca{Ca~{\sevenrm II}}
\def\Mgb{Mg~{\sevenrm Ib}}
\def\CIII{C~{\sevenrm III}]}
\def\Mg{Mg~{\sevenrm II}}
\def\NeIII{[Ne~{\sevenrm III}]}
\def\NeV{[Ne~{\sevenrm V}]}
\def\OII{[O~{\sevenrm II}]}
\def\OIII{[O~{\sevenrm III}]}
\def\Ha{H{\sevenrm $\alpha$}}
\def\Hb{H{\sevenrm $\beta$}}
\def\Hg{H{\sevenrm $\gamma$}}
\def\Hd{H{\sevenrm $\delta$}}

\newdimen\digitwidth
\setbox0=\hbox{-.}
\digitwidth=\wd0
\catcode `@=\active
\def@{\kern\digitwidth}

\date{Accepted ~~. Received ~~; in original form ~~}

\pagerange{\pageref{firstpage}--\pageref{lastpage}} \pubyear{2008}

\maketitle

\label{firstpage}

\begin{abstract}

We present deep radio images at 1.4 GHz of a large and complete sample
of BL Lacertae objects (BL Lacs) selected from the Deep X-ray Radio
Blazar Survey (DXRBS). We have observed 24 northern ($\delta \ga
-30^{\circ}$) sources with the Very Large Array (VLA) in both its A
and C configurations and 15 southern sources with the Australia
Telescope Compact Array (ATCA) in its largest configuration. We find
that in the DXRBS, as in the 1-Jy survey, which has a radio flux limit
roughly ten times higher than the DXRBS, a considerable number (about
a third) of BL Lacs can be identified with the relativistically beamed
counterparts of Fanaroff-Riley type II (FR II) radio galaxies. We
attribute the existence of FR II-BL Lacs, which is not accounted for
by current unified schemes, to an inconsistency in our classification
scheme for radio-loud active galactic nuclei (AGN). Taking the
extended radio power as a suitable measure of intrinsic jet power, we
find similar average values for low- (LBL) and high-energy peaked BL
Lacs (HBL), contrary to the predictions of the blazar sequence.

\end{abstract}

\begin{keywords}
BL Lacertae objects: general - galaxies: active - radio continuum: galaxies
\end{keywords}

\section{Introduction}

The basic nature of BL Lacertae objects (BL Lacs) is believed to be
understood within the unified schemes for radio-loud active galactic
nuclei (AGN). These sources are radio galaxies, which have their
relativistic jets oriented close to the observer's line of sight. For
objects with such a preferred jet alignment, which are commonly
referred to as 'blazars', the relativistic beaming effect can explain
most of the observed properties, such as, e.g., non-thermal continuum
emission from radio up to $\gamma$-ray frequencies, high core
luminosities, core-dominated radio morphologies, irregular and rapid
variability, and strong radio and optical polarization (see review by
\citet{Urry95} and references therein).

Current unified schemes identify the parent population of BL Lacs with
the low-luminosity Fanaroff-Riley type I \citep[FR I;][]{Fan74} radio
galaxies, whereas the high-luminosity Fanaroff-Riley type II (FR II)
radio galaxies are assumed to appear as radio quasars when their jets
are viewed at relatively small angles. However, in recent years
evidence has accumulated that some BL Lacs might in fact be beamed FR
II radio galaxies. Their extended radio powers are higher than those
of known FR Is and their extended radio morphologies are consistent
with those of FR IIs once orientation effects are accounted for
\citep[][]{Kol92, Cas99, Rec01}.

If indeed a considerable number of FR IIs are part of the BL Lac
parent population, it will have important implications. Estimates of
relativistic beaming parameters such as, e.g., the jet bulk Lorentz
factor or viewing angle, which are usually derived from a comparison
of the radio luminosity function \citep[e.g.,][]{Urry91, Urry95} or
the distribution of radio core-dominance values \citep[e.g.,][]{Kol92,
Per93} of the parent population with that of the corresponding blazar
class, will have to be revisited. Furthermore, since, by definition,
the broad emission line region, which is assumed to trace directly the
accretion disk power, is absent or only very weak in BL Lacs, the
existence of a considerable number of FR II-like jet powers among this
blazar class could mean that accretion disk and jet luminosities are
not as closely linked as current jet formation models suggest
\citep[e.g.,][]{Bla82, Raw91, Mar03}.

A further implication will concern the so-called 'blazar sequence'
\citep{Fos98, Ghi98}. This model posits that Compton-cooling
determines the frequency of the jet synchrotron emission peak, in
particular that the higher the (intrinsic) jet power, the stronger the
cooling, and the lower the synchrotron emission peak
frequency. Therefore, finding in particular high-energy peaked BL Lacs
with FR II-like extended radio powers could present a severe challenge
for this model \citep[see also][]{Pad07}.

Our current knowledge of the radio properties of BL Lacs is based
mainly on deep radio observations of two complete samples selected at
widely different frequencies, namely, the radio-selected 1-Jy
\citep{Sti91, Rec01} and the X-ray-selected {\it Einstein} Medium
Sensitivity Survey \citep[EMSS;][]{Mor91, Rec00} BL Lac
samples. However, these two surveys sample the extreme ends of the
radio flux distribution of BL Lacs and, therefore, have presented a
strongly biased view of BL Lac physics. In order to rectify this
situation we have obtained deep radio images of a complete sample of
BL Lacs with intermediate radio properties.

The paper is structured as follows. In Section 2 we discuss the
selection of the BL Lac sample, for which the radio images have been
obtained and analyzed as detailed in Section 3. In Section 4 we
address the question of the parent population of BL Lacs. In Section 5
we investigate if some BL Lacs with featureless optical spectra are at
high redshifts rather than strongly relativistically beamed. The radio
properties of low- and high-energy peaked BL Lacs are compared in
Section 6. Finally, Section 7 summarizes of our main results and
presents our conclusions. For consistency with previous work we have
assumed throughout this paper cosmological parameters $H_0 = 50$ km
s$^{-1}$ Mpc$^{-1}$ and $q_0 = 0$. Energy spectral indices have been
defined as $f_\nu \propto \nu^{-\alpha}$.

\section{The Sample}


\begin{table*}
\caption{\label{general} 
General Properties of the Observed Sample and Log of Observations}
\begin{tabular}{lcccccrrrclll}
\hline
Object Name & CS & class & z & z & z & $f_{\rm NVSS}$ & $\alpha_{\rm r}$ & $\alpha_{\rm r}$ & 
$f_{\rm 1keV}$ & observation & Array & program \\ 
&&&& type & flag & [mJy] & (1.4-5) & (4.8-8.6) & [$\mu$Jy] & date && \\
(1) & (2) & (3) & (4) & (5) & (6) & (7) & (8) & (9) & (10) & (11) & (12) & (13) \\
\hline
WGA J0032$-$2849 & Y & WL & 0.324 & abs & c & 159.9 & $ $0.08 & $ $0.14 & 0.047 & 2004 Sep 30 & VLA A   & AL627 \\
                 &   &    &       &     &   &       &         &         &       & 2002 Sep 30 & VLA C   & AL578 \\
WGA J0040$-$2340 &   &    & 0.213 &  ?  & ? &  52.9 & $ $0.09 & $ $0.33 & 0.009 & 2004 Oct 7  & VLA A   & AL627 \\
                 &   &    &       &     &   &       &         &         &       & 2002 Oct 2  & VLA C   & AL578 \\
WGA J0043$-$2638 & Y &    & 1.003 & em  & c &  77.8 & $-$0.03 & $ $0.12 & 0.062 & 2004 Oct 7  & VLA A   & AL627 \\
                 &   &    &       &     &   &       &         &         &       & 2002 Sep 30 & VLA C   & AL578 \\
WGA J0100$-$3337 & Y &    & 0.875 & em  & p &  70.4 & $-$0.42 & $-$0.21 & 0.010 & 2004 Jan 26 & ATCA 6B & C1010 \\
WGA J0245$+$1047 & Y & SL & 0.070 & em  & c & 417.4 & $ $0.55 &         & 0.262 & 2002 Feb 6  & VLA A   & AL564 \\
                 &   &    &       &     &   &       &         &         &       & 2005 Sep 12 & VLA C   & AL652 \\
WGA J0313$+$4115 &   & WL & 0.029 & em  & c &  40.6 & $-$0.04 &         & 0.112 & 2002 Jan 22 & VLA A   & AL564 \\
                 &   &    &       &     &   &       &         &         &       & 2005 Sep 12 & VLA C   & AL652 \\
WGA J0428$-$3805 & Y & WL & 0.150 & abs & c &$>$47.9& $ $0.58 & $ $1.98 & 0.004 & 2002 May 23 & ATCA 6A & C1010 \\
                 &   &    &       &     &   &       &         &         &       & 2004 Jan 26 & ATCA 6B & C1010 \\
WGA J0431$+$1731 & Y & WL & 0.143 & abs & t & 374.3 & $ $0.30 &         & 0.035 & 2002 Jan 22 & VLA A   & AL564 \\
                 &   &    &       &     &   &       &         &         &       & 2002 Oct 11 & VLA C   & AL578 \\
WGA J0449$-$4349 &   & WL &$>$0.176&    & l &       &         & $ $0.15 & 0.477 & 2002 May 23 & ATCA 6A & C1010 \\
                 &   &    &       &     &   &       &         &         &       & 2004 Jan 26 & ATCA 6B & C1010 \\
WGA J0528$-$5820 & Y & WL & 0.254 & em  & c &       &         & $ $0.70 & 0.031 & 2002 May 22 & ATCA 6A & C1010 \\
                 &   &    &       &     &   &       &         &         &       & 2004 Jan 26 & ATCA 6B & C1010 \\
WGA J0533$-$4632 & Y & WL & 0.332 & abs & c &       &         & $ $0.60 & 0.022 & 2004 Jan 26 & ATCA 6B & C1010 \\
WGA J0558$+$5328 & Y &    & 0.036 & em  & c & 386.4 & $ $0.59 &         & 0.040 & 2002 Feb 10 & VLA A   & AL564 \\
                 &   &    &       &     &   &       &         &         &       & 2002 Oct 11 & VLA C   & AL578 \\
WGA J0624$-$3229 & Y & WL & 0.249 & abs & c &  42.2 & $-$0.53 &         & 0.210 & 2006 Feb 6  & VLA A   & AP501 \\
                 &   &    &       &     &   &       &         &         &       & 2002 Oct 5  & VLA C   & AL578 \\
WGA J0847$+$1133 &   & WL & 0.198 & abs & c &  32.8 & $ $0.03 &         & 3.954 & 2006 May 13 & VLA A   & AL664 \\
                 &   &    &       &     &   &       &         &         &       & 2005 Sep 12 & VLA C   & AL652 \\
WGA J0940$+$2603 & Y & WL &$>$0.567&    & l & 462.2 & $ $0.05 &         & 0.082 & 2002 Apr 22 & VLA A   & AL564 \\
                 &   &    &       &     &   &       &         &         &       & 2005 Sep 12 & VLA C   & AL652 \\
WGA J1204$-$0710 & Y & WL & 0.183 & abs & c & 167.7 & $ $0.22 & $ $0.25 & 0.070 & 2006 May 13 & VLA A   & AL664 \\
                 &   &    &       &     &   &       &         &         &       & 2005 Sep 12 & VLA C   & AL652 \\
WGA J1231$+$2848 &   & WL &$>$0.878&    & l & 140.5 & $ $0.52 &         & 0.150 & 2002 Feb 26 & VLA A   & AL564 \\
                 &   &    &       &     &   &       &         &         &       & 2005 Sep 12 & VLA C   & AL652 \\
WGA J1311$-$0521 &   & WL & 0.160 & abs & c &  69.8 & $ $0.36 & $ $0.23 & 0.040 & 2002 Feb 26 & VLA A   & AL564 \\
                 &   &    &       &     &   &       &         &         &       & 2005 Sep 12 & VLA C   & AL652 \\
WGA J1320$+$0140 &   &    & 1.235 & em  & c & 670.7 & $ $0.01 &         & 0.036 & 2002 Feb 4  & VLA A   & AL564 \\
                 &   &    &       &     &   &       &         &         &       & 2005 Sep 12 & VLA C   & AL652 \\
WGA J1539$-$0658 & Y & WL &$>$0.525&    & l &  46.8 & $-$0.27 & $ $0.47 & 0.018 & 2002 Feb 4  & VLA A   & AL564 \\
                 &   &    &       &     &   &       &         &         &       & 2005 Sep 12 & VLA C   & AL652 \\
WGA J1744$-$0517 & Y & WL & 0.310 & abs & t & 196.1 & $ $0.69 &         & 0.024 & 2006 May 13 & VLA A   & AL664 \\
                 &   &    &       &     &   &       &         &         &       & 2005 Sep 12 & VLA C   & AL652 \\
WGA J1808$+$0546 & Y & WL & 0.180 & em  & c & 176.3 & $ $0.01 &         & 0.014 & 2006 May 13 & VLA A   & AL664 \\
WGA J1834$-$5856 & Y & WL &$>$0.596&    & l &       &         & $ $0.00 & 0.102 & 2004 Jan 26 & ATCA 6B & C1010 \\
WGA J1834$-$5948 & Y & WL & 0.435 & abs & t &       &         & $ $0.63 & 0.062 & 2004 Jan 26 & ATCA 6B & C1010 \\
WGA J1840$+$5452 & Y & SL & 0.646 & em  & c & 207.3 & $ $0.39 &         & 0.179 & 2002 Apr 7  & VLA A   & AL564 \\
                 &   &    &       &     &   &       &         &         &       & 2005 Sep 12 & VLA C   & AL652 \\
WGA J1843$-$7430 &   & WL & 0.166 & abs & t &       &         & $ $0.39 & 0.005 & 2004 Jan 26 & ATCA 6B & C1010 \\
WGA J1936$-$4719 & Y & WL & 0.264 & abs & p &       &         & $-$0.11 & 0.936 & 2004 Jan 25 & ATCA 6B & C1010 \\
WGA J2258$-$5525 & Y & WL & 0.479 & em  & p &       &         & $ $0.10 & 0.201 & 2004 Jan 25 & ATCA 6B & C1010 \\
WGA J2322$-$4221 &   & WL & 0.089 & abs & c &       &         & $ $0.02 & 0.019 & 2004 Jan 25 & ATCA 6B & C1010 \\
WGA J2330$-$3724 & Y & WL & 0.279 & em  & c & 304.5 & $ $0.19 & $-$0.18 & 0.038 & 2004 Jan 25 & ATCA 6B & C1010 \\
EXO 0556$-$3838  & Y & WL & 0.302 & abs & c & 104.6 & $ $0.29 &         & 3.005 & 2002 May 22 & ATCA 6A & C1010 \\
EXO 1811$+$3143  & Y &    & 0.117 &  ?  & ? & 191.4 & $ $0.17 &         & 0.090 & 2002 Apr 7  & VLA A   & AL564 \\
                 &   &    &       &     &   &       &         &         &       & 2005 Sep 12 & VLA C   & AL652 \\
MH 2136$-$428    & Y & WL &$>$0.262&    & l &       &         & $-$0.13 & 0.113 & 2002 Feb 4  & ATCA 6B & C1009$^\star$\\
PKS 2316$-$423   &   & WL & 0.055 & em  & c &       &         & $ $0.43 & 0.337 & 2004 Jan 25 & ATCA 6B & C1010 \\
PMN J0630$-$24   & Y & WL & 1.238 & abs & c & 105.4 & $-$0.16 &         & 0.322 & 2002 Feb 10 & VLA A   & AL564 \\
                 &   &    &       &     &   &       &         &         &       & 2002 Sep 20 & VLA C   & AL578 \\
RX J09168$+$523  &   & WL & 0.190 & abs & c & 137.8 & $ $0.58 &         & 0.360 & 2002 Apr 22 & VLA A   & AL564 \\
                 &   &    &       &     &   &       &         &         &       & 2005 Sep 12 & VLA C   & AL652 \\
TEX 0836$+$182   & Y & WL &$>$0.465&    & l & 409.9 & $ $0.17 &         & 0.080 & 2006 May 13 & VLA A   & AL664 \\
                 &   &    &       &     &   &       &         &         &       & 2002 Oct 18 & VLA C   & AL578 \\
\hline		 
\end{tabular}
\end{table*}

\begin{table*}
\contcaption{}
\begin{tabular}{lcccccrrrclll}
\hline
Object Name & CS & class & z & z & z & $f_{\rm NVSS}$ & $\alpha_{\rm r}$ & $\alpha_{\rm r}$ & 
$f_{\rm 1keV}$ & observation & Array & program \\ 
&&&& type & flag & [mJy] & (1.4-5) & (4.8-8.6) & [$\mu$Jy] & date && \\
(1) & (2) & (3) & (4) & (5) & (6) & (7) & (8) & (9) & (10) & (11) & (12) & (13) \\
\hline
WGA 1012$+$06    & Y &    & 0.727 & em  & c & 535.2 & $ $0.45 &         & 0.055 & 2006 May 13 & VLA A   & AL664 \\
                 &   &    &       &     &   &       &         &         &       & 2005 Sep 12 & VLA C   & AL652 \\
WGA 1202$+$44    &   & WL & 0.297 & abs & c & 105.0 & $ $1.43 &         & 0.112 & 2006 May 13 & VLA A   & AL664 \\
                 &   &    &       &     &   &       &         &         &       & 2005 Sep 12 & VLA C   & AL652 \\
\hline
\end{tabular}

\medskip

\parbox[]{14.5cm}{The columns are: (1) object name; (2) 'Y' if part of
  the complete sample; (3) classification following \citet{L04}, where
  WL: weak-lined radio-loud AGN, SL: strong-lined radio-loud AGN; (4)
  redshift; (5) type of redshift, where abs: based on absorption
  lines, em: based on emission lines, ?: optical spectrum not
  published; (6) reliability of redshift, where c: certain (based on
  two or more lines), p: possible (based on one emission line or two
  or more weak absorption lines), t: tentative (based on low S/N
  optical spectrum), ?: optical spectrum not published, and l: lower
  limit estimated from the optical $V$ magnitude following
  \citet{Pir07}; (7) total radio flux at 1.4 GHz from the NVSS; (8)
  radio spectral index between 1.4 and 5 GHz, calculated from the sum
  of the fluxes of all NVSS sources within a $3'$ radius
  (corresponding roughly to the beam size of the GB6 survey) and the
  total flux from the GB6 and PMN surveys for northern and southern
  sources, respectively; (9) radio spectral index between 4.8 and 8.6
  GHz from ATCA snapshot observations; (10) unabsorbed {\it ROSAT}
  X-ray flux at 1 keV, calculated using an X-ray spectral index
  derived from hardness ratios; (11) observation date; (12) telescope
  configuration; and (13) program number.}

\medskip

\parbox[]{14.5cm}{$^\star$ observed at 4.8 GHz}

\end{table*}


\begin{table*}
\caption{\label{generalpub} 
General Properties of the Sources with Published 1.4 GHz Observations}
\begin{tabular}{lcccccrrcll}
\hline
Object Name & CS & class & z & z & z & $f_{\rm NVSS}$ & $\alpha_{\rm r}$ & 
$f_{\rm 1keV}$ & Array & Reference \\ 
&&&& type & flag & [mJy] & (1.4-5) & [$\mu$Jy] && \\
(1) & (2) & (3) & (4) & (5) & (6) & (7) & (8) & (9) & (10) & (11) \\
\hline
B2 1147$+$245    & Y & WL &$>$0.292&    & l &  797.7 & $-$0.02 & 0.022 & VLA A   & \citet{Ant85b} \\
ON 325           & Y &    & 0.237 &  ?  & ? &  571.6 & $ $0.09 & 0.473 & VLA A   & \citet{Ant85b} \\
1ES 1212$+$078   &   & WL & 0.130 & abs & c &  136.6 & $-$0.04 & 0.945 & VLA A   & \citet{Gir04}  \\
3C 66A           & Y &    & 0.444 &  ?  & ? & 2303.1 & $ $0.70 & 1.301 & VLA BnA & \citet{Ant85b} \\
4C $+$55.17      & Y & SL & 0.900 & em  & c & 3079.2 & $ $0.33 & 0.034 & VLA A   & \citet{Mur93}  \\
\hline
\end{tabular}

\medskip

\parbox[]{14.5cm}{The columns are: (1) - (8) as in Table
  \ref{general}; (9) unabsorbed {\it ROSAT} X-ray flux at 1 keV,
  calculated using an X-ray spectral index derived from hardness
  ratios; (10) telescope configuration; and (11) reference.}

\end{table*}

We have selected for radio imaging BL Lacs from the Deep X-ray Radio
Blazar Survey \citep[DXRBS;][]{Per98, L01, P07}. In short, the DXRBS
takes advantage of the fact that blazars are relatively strong X-ray
and radio emitters and, by definition, have a flat radio spectrum. In
this spirit it is the result of a cross-correlation between the {\it
ROSAT} database of serendipitous X-ray sources WGACAT95 \citep[second
revision:][]{Whi95} and a number of publicly available radio catalogs
(the 20 cm and 6 cm Green Bank survey catalogs NORTH20CM and GB6
\citep{Gre96, White92} for the northern hemisphere and the 6 cm
Parkes-MIT-NRAO \citep[PMN;][]{Gri93} catalog for the southern
hemisphere). All sources with radio spectral index $\alpha_{\rm r}
\leq 0.7$ and off the Galactic plane ($|b| > 10^{\circ}$) were
selected as blazar candidates. Radio spectral indices for southern
sources, for which a survey at a frequency different from that of the
PMN was not available when the project was started, were obtained from
a snapshot survey conducted with the Australia Telescope Compact Array
(ATCA) at 3.6 and 6 cm.

The DXRBS complete sample is $\sim 95$ per cent optically identified
\citep{P07}. The total sample (i.e., the complete sample plus a
subsample of 'low priority' sources) contains over 300 blazars, of
which the large majority are radio quasars and 44 are BL Lacs. The
general properties of the DXRBS BL Lacs observed by us and with
published radio data useful for our purpose are listed in Tables
\ref{general} and \ref{generalpub}, respectively. Two DXRBS blazar
candidates previously classified as BL Lacs are now considered
optically unidentified based on radio images obtained as part of this
program. For WGA J0023$+$0417, a Very Large Array (VLA) A
configuration map shows the NRAO/VLA Sky Survey (NVSS) radio source,
whose position was used to pinpoint the optical candidate, to be
composed of two sources $\sim 50''$ apart. For WGA J0816$-$0736, a VLA
C array map shows a radio source with an extended core-jet
morphology. The core is $\sim 95''$ away from the observed optical
candidate.

\subsection{The Classification} \label{class}

The first criteria for the classification of a radio-loud AGN as a BL
Lac, namely, a compact radio morphology and a completely featureless
optical spectrum \citep{Str72}, have been modified several times over
the years and most recently by \citet{L02, L04}. The classification
problem is twofold. On the one hand, BL Lacs are blazars, i.e., they
are strongly relativistically beamed, and need to be efficiently
separated from radio galaxies. Following \citet{Sto91} and
\citet{Marcha96}, this is accomplished in the DXRBS using the value of
the Ca H\&K break, a discontinuity in the host galaxy spectrum, which
is defined as $C = (f_+ - f_-)/f_+$, where $f_-$ and $f_+$ are the
fluxes in the rest-frame wavelength regions $3750-3950$~\AA~and
$4050-4250$~\AA, respectively. DXRBS BL Lacs are defined to have
$C<0.4$, since \citet{L02} showed that above this value sources become
increasingly lobe-dominated (see their Fig. 6).

On the other hand, blazars can have both {\it intrinsically} weak and
strong emission line regions and a separation between BL Lacs and
radio quasars is meant to reflect this. Initially, the DXRBS used the
scheme proposed by \citet{Marcha96} to classify blazars as BL
Lacs. However, as discussed by \citet{L04}, since this scheme imposes
a limit on the equivalent width of the strongest, observed emission
line (i.e., narrow or broad and independent of redshift), it is rather
arbitrary. Following \citet{L04}, the DXRBS now uses the rest-frame
equivalent width plane of the narrow emission lines \OII~$\lambda
3727$ and \OIII~$\lambda 5007$ (see their Fig. 4) to separate BL Lacs
(i.e., weak-lined radio-loud AGN) and radio quasars (i.e.,
strong-lined radio-loud AGN). In Tables \ref{general} and
\ref{generalpub}, column (3), we list the classification following
\citet{L04}. For 9/44 sources we currently cannot apply their scheme
since either we have no optical spectrum (4 sources) or the available
spectrum does not cover the wavelength of \OIII~(5 sources). We note
that whereas the large majority of the DXRBS BL Lacs classified in
this way are weak-lined radio-loud AGN, three sources are strong-lined
radio-loud AGN, i.e., they might be more similar to radio quasars than
to BL Lacs (see also Section \ref{incon}).

\subsection{The Redshift Determination} \label{red}

The weakness of both emission and absorption features in the optical
spectra of BL Lacs renders the determination of their redshift a
difficult task and in some cases an even impossible one. Nevertheless,
based on new optical spectroscopy, which will be presented in a future
paper, and a new method introduced by \citet{Pir07} all DXRBS BL Lacs
now have a redshift or a lower limit thereof (Tables \ref{general} and
\ref{generalpub}, columns (3)-(5)). Emission lines are detected in
14/40 sources, for which an optical spectrum is available. We regard
the redshifts of 25/32 sources (or 78 per cent), for which an optical
spectrum is available and emission and/or absorption features are
detected, as firm. 

The optical spectrum of 8/40 BL Lacs appears featureless. For these
sources we have determined lower limits on the redshift following
\citet{Pir07}. Their method uses the fact that BL Lacs are hosted by
ellipticals of almost constant luminosity \citep[e.g.,][]{Urry00},
i.e., by standard candles. Then, assuming a plausible lower limit for
the jet/galaxy ratio of a featureless BL Lac one can derive a lower
limit on the redshift from the observed optical $V$ magnitude.

\section{Observations and Data Reduction}


\begin{figure*}
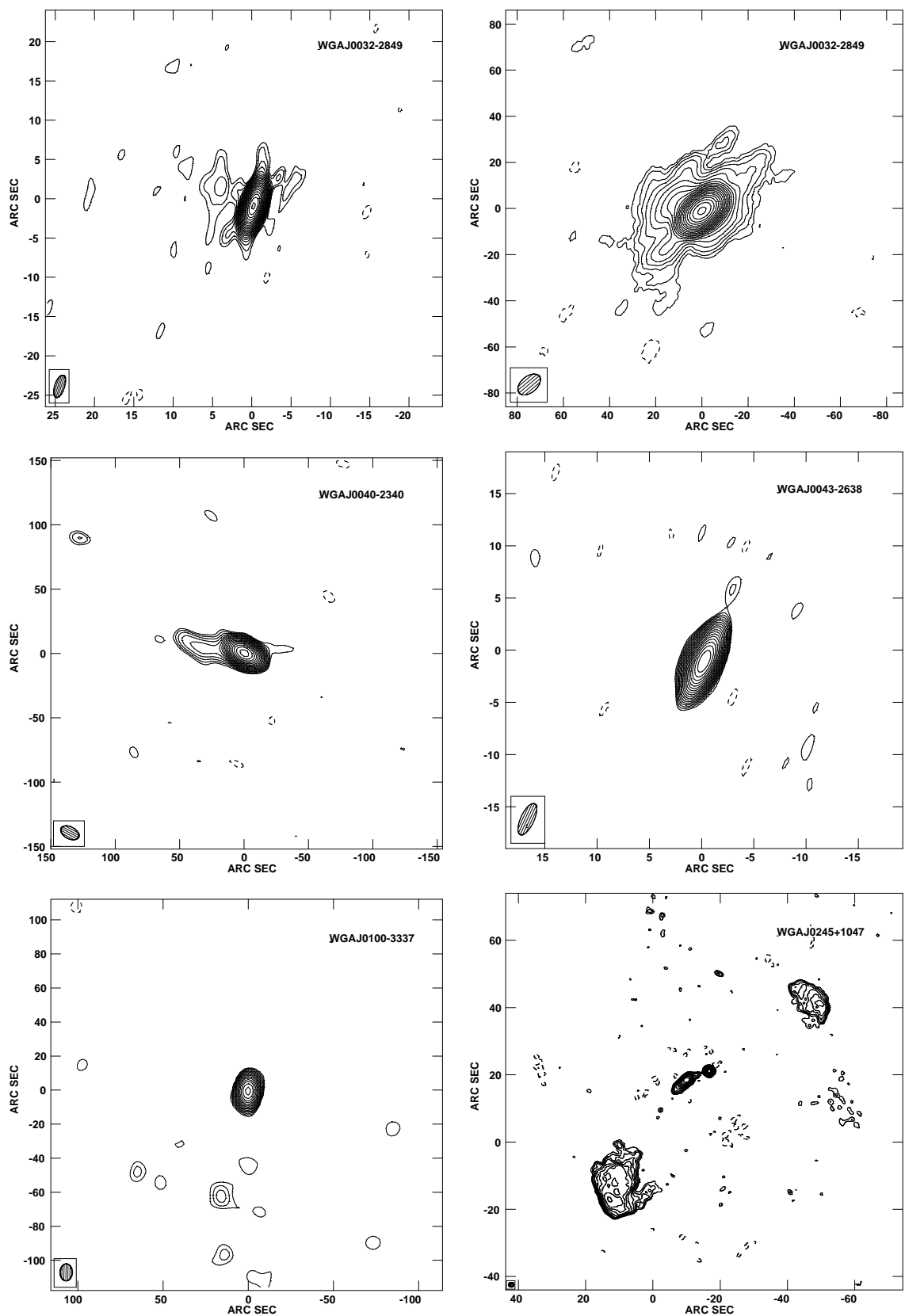

\centerline{
\includegraphics[scale=0.4]{A_WGAJ0032-2849.PS}
\includegraphics[scale=0.4]{AC_WGAJ0032-2849.PS}
}
\centerline{
\includegraphics[scale=0.4]{C_WGAJ0040-2340.PS}
\includegraphics[scale=0.4]{A_WGAJ0043-2638.PS}
}
\centerline{
\includegraphics[scale=0.4]{6B_WGAJ0100-3337.PS}
\includegraphics[scale=0.4]{A_WGAJ0245+1047.PS}
}
\caption{\label{maps1} (a) WGA J0032$-$2849, VLA A. Image rms is 0.07
mJy/beam. Image peak is 151.8 mJy/beam. (b) WGA J0032$-$2849, VLA
A$+$C. Image rms is 0.06 mJy/beam. Image peak is 144.6 mJy/beam. (c)
WGA J0040$-$2340, VLA C. Image rms is 0.05 mJy/beam. Image peak is
45.9 mJy/beam. (d) WGA J0043$-$2638, VLA A. Image rms is 0.05
mJy/beam. Image peak is 76.4 mJy/beam. (e) WGA J0100$-$3337, ATCA
6B. Image rms is 0.20 mJy/beam. Image peak is 90.1 mJy/beam. (f) WGA
J0245$+$1047, VLA A. Image rms is 0.04 mJy/beam. Image peak is 28.6
mJy/beam. In all images contours start at 3 times the rms and positive
values are spaced by factors of $\sqrt{2}$ up to the image peak.}
\end{figure*}


\begin{figure*}
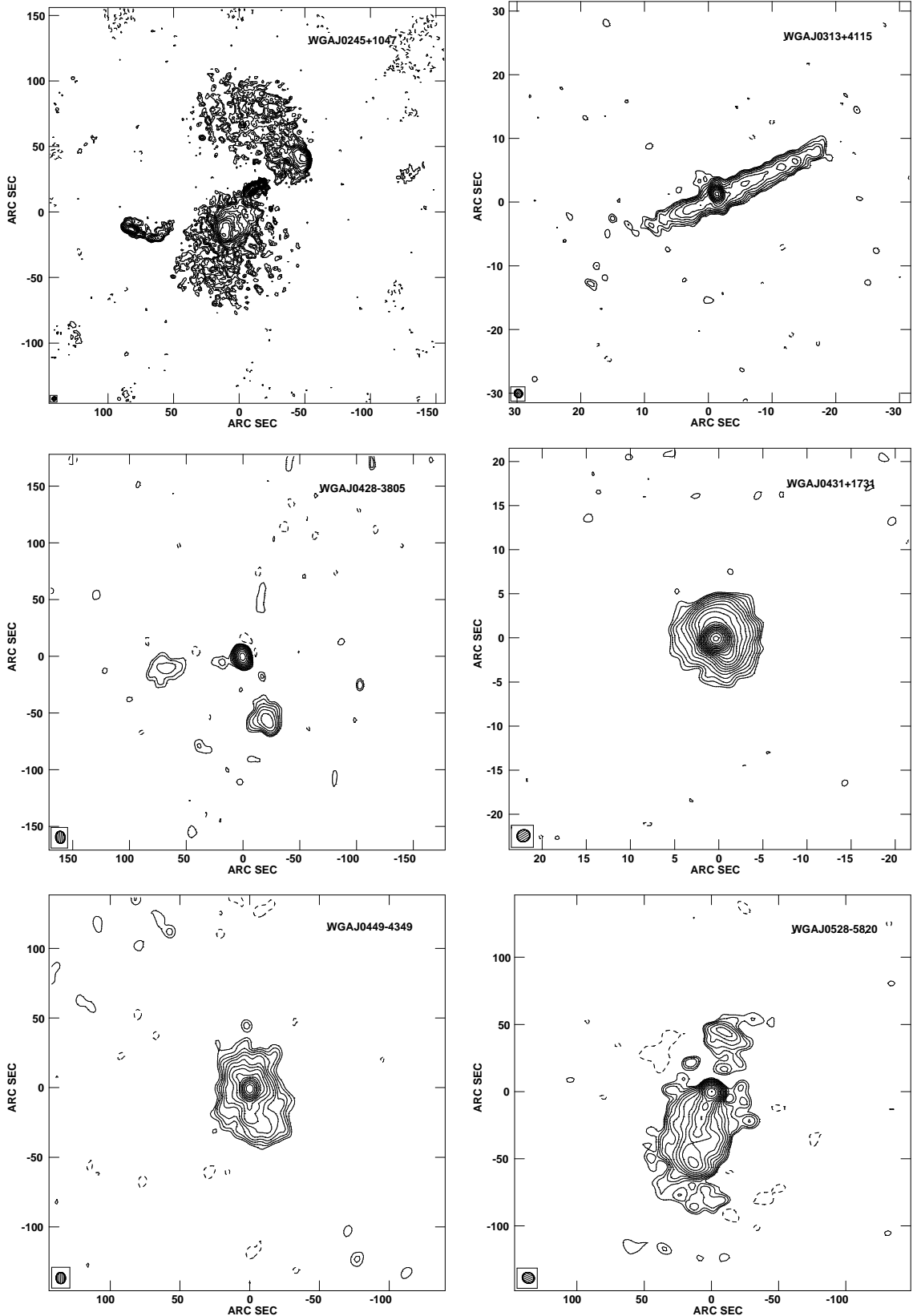

\centerline{
\includegraphics[scale=0.4]{AC_WGAJ0245+1047.PS}
\includegraphics[scale=0.4]{A_WGAJ0313+4115.PS}
}
\centerline{
\includegraphics[scale=0.4]{6AB_WGAJ0428-3805.PS}
\includegraphics[scale=0.4]{A_WGAJ0431+1731.PS}
}
\centerline{
\includegraphics[scale=0.4]{6AB_WGAJ0449-4349.PS}
\includegraphics[scale=0.4]{6AB_WGAJ0528-5820.PS}
}
\caption{\label{maps2} (a) WGA J0245$+$1047, VLA A$+$C. Image rms is
  0.04 mJy/beam. Image peak is 30.5 mJy/beam. (b) WGA J0313$+$4115,
  VLA A. Image rms is 0.17 mJy/beam. Image peak is 54.8 mJy/beam. (c)
  WGA J0428$-$3805, ATCA 6A$+$6B. Image rms is 0.24 mJy/beam. Image
  peak is 476.5 mJy/beam. (d) WGA J0431$+$1731, VLA A. Image rms is
  0.08 mJy/beam. Image peak is 262.6 mJy/beam. (e) WGA J0449$-$4349,
  ATCA 6A$+$6B. Image rms is 0.18 mJy/beam. Image peak is 99.3
  mJy/beam. (f) WGA J0528$-$5820, ATCA 6A$+$6B. Image rms is 0.10
  mJy/beam. Image peak is 26.1 mJy/beam. Contours and positive values
  as in Fig. \ref{maps1}.}
\end{figure*}


\begin{figure*}
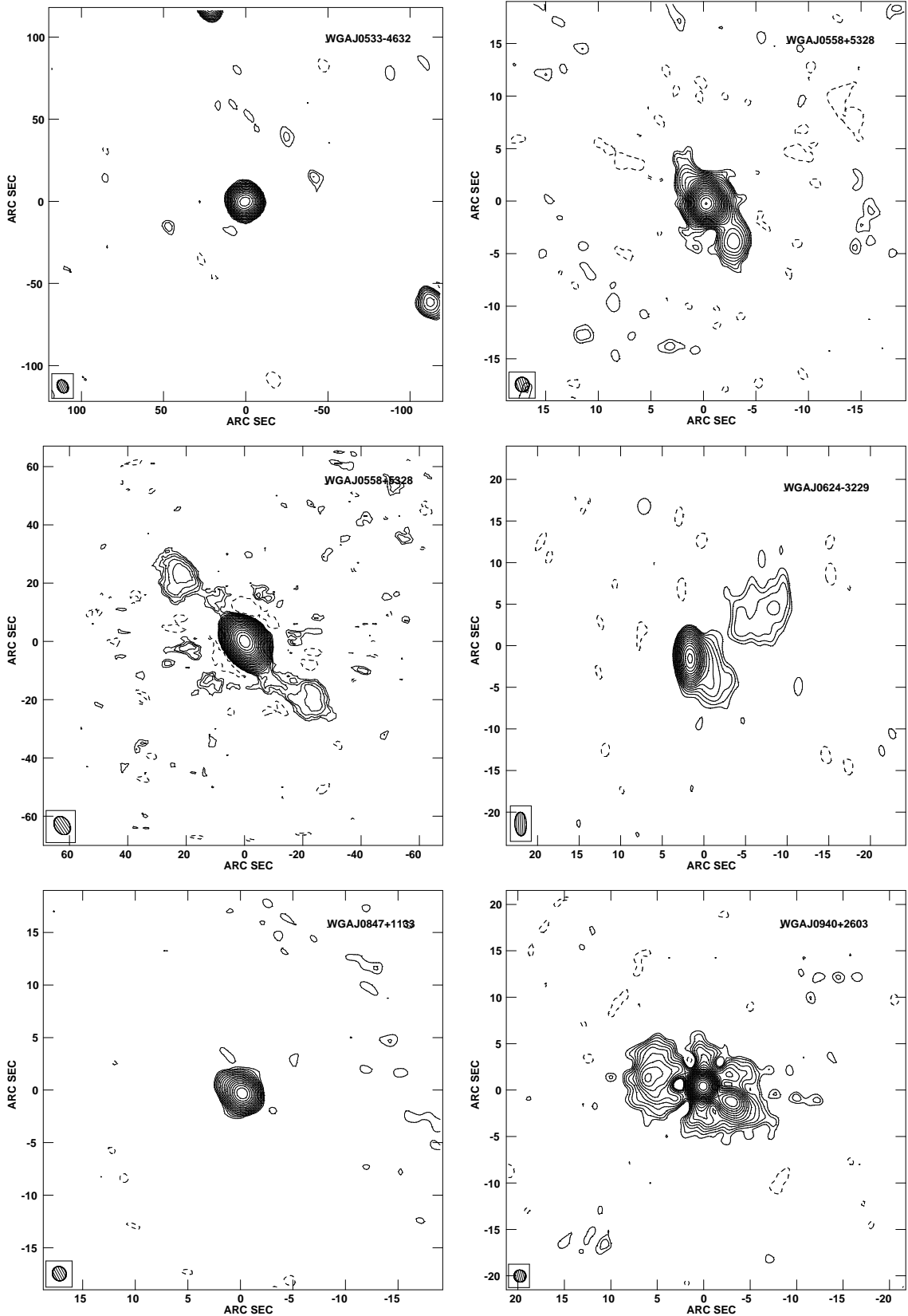

\centerline{
\includegraphics[scale=0.4]{6B_WGAJ0533-4632.PS}
\includegraphics[scale=0.4]{A_WGAJ0558+5328.PS}
}
\centerline{
\includegraphics[scale=0.4]{AC_WGAJ0558+5328.PS}
\includegraphics[scale=0.4]{A_WGAJ0624-3229.PS}
}
\centerline{
\includegraphics[scale=0.4]{A_WGAJ0847+1133.PS}
\includegraphics[scale=0.4]{A_WGAJ0940+2603.PS}
}
\caption{\label{maps3} (a) WGA J0533$-$4632, ATCA 6B. Image rms is
  0.11 mJy/beam. Image peak is 108.2 mJy/beam. (b) WGA J0558$+$5328,
  VLA A. Image rms is 0.07 mJy/beam. Image peak is 296.9 mJy/beam. (c)
  WGA J0558$+$5328, VLA A$+$C. Image rms is 0.06 mJy/beam. Image peak
  is 355.1 mJy/beam. (d) WGA J0624$-$3229, VLA A. Image rms is 0.06
  mJy/beam. Image peak is 34.8 mJy/beam. (e) WGA J0847$+$1133, VLA
  A. Image rms is 0.05 mJy/beam. Image peak is 26.9 mJy/beam. (f) WGA
  J0940$+$2603, VLA A. Image rms is 0.06 mJy/beam. Image peak is 371.7
  mJy/beam. Contours and positive values as in Fig. \ref{maps1}.}
\end{figure*}


\begin{figure*}
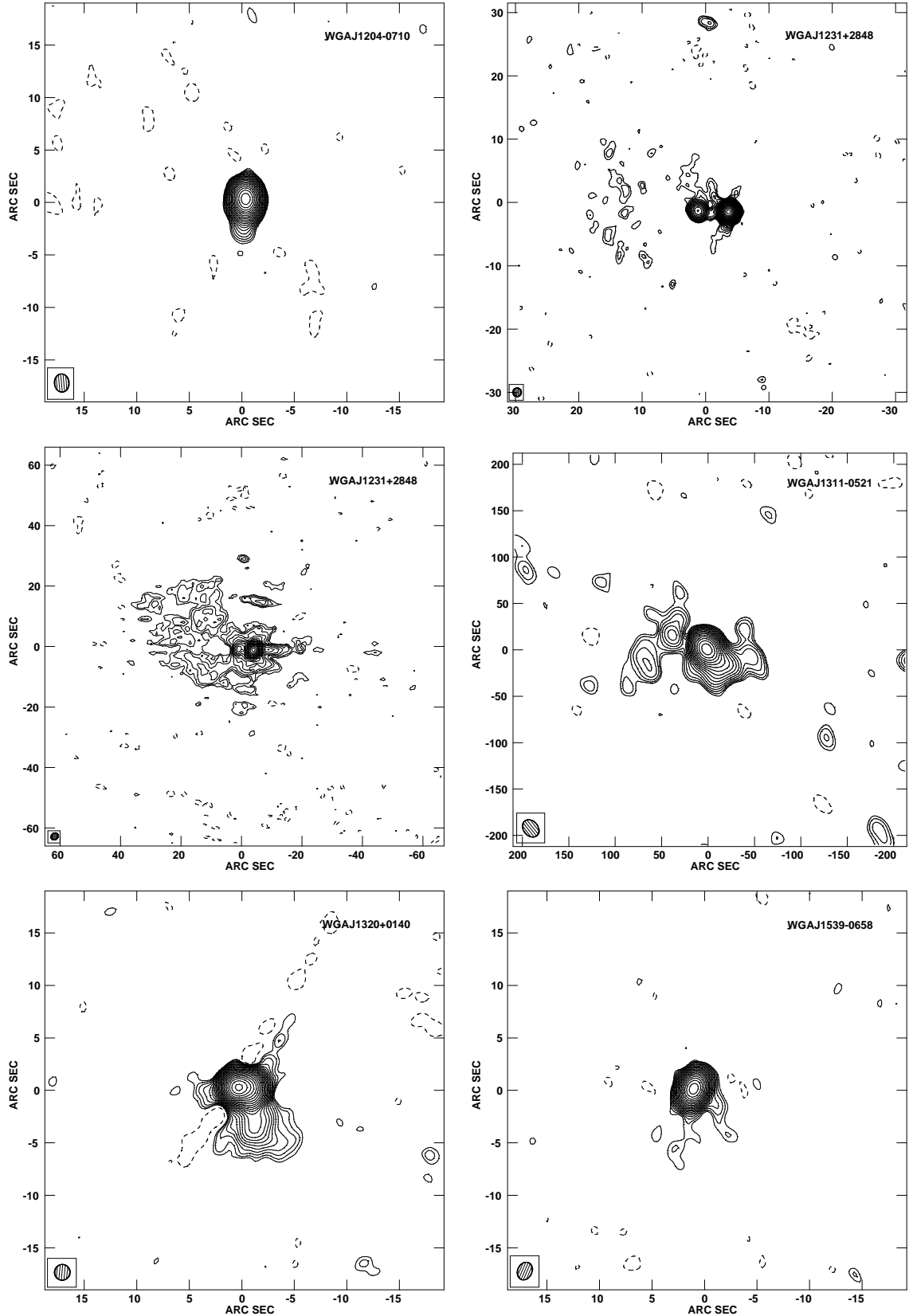

\centerline{
\includegraphics[scale=0.4]{A_WGAJ1204-0710.PS}
\includegraphics[scale=0.4]{A_WGAJ1231+2848.PS}
}
\centerline{
\includegraphics[scale=0.4]{AC_WGAJ1231+2848.PS}
\includegraphics[scale=0.4]{C_WGAJ1311-0521.PS}
}
\centerline{
\includegraphics[scale=0.4]{A_WGAJ1320+0140.PS}
\includegraphics[scale=0.4]{A_WGAJ1539-0658.PS}
}
\caption{\label{maps4} (a) WGA J1204$-$0710, VLA A. Image rms is 0.08
  mJy/beam. Image peak is 148.1 mJy/beam. (b) WGA J1231$+$2848, VLA
  A. Image rms is 0.02 mJy/beam. Image peak is 85.6 mJy/beam. (c) WGA
  J1231$+$2848, VLA A$+$C. Image rms is 0.04 mJy/beam. Image peak is
  90.5 mJy/beam. (d) WGA J1311$-$0521, VLA C. Image rms is 0.07
  mJy/beam. Image peak is 79.1 mJy/beam. (e) WGA J1320$+$0140, VLA
  A. Image rms is 0.07 mJy/beam. Image peak is 496.5 mJy/beam. (f) WGA
  J1539$-$0658, VLA A. Image rms is 0.02 mJy/beam. Image peak is 43.5
  mJy/beam. Contours and positive values as in Fig. \ref{maps1}.}
\end{figure*}


\begin{figure*}
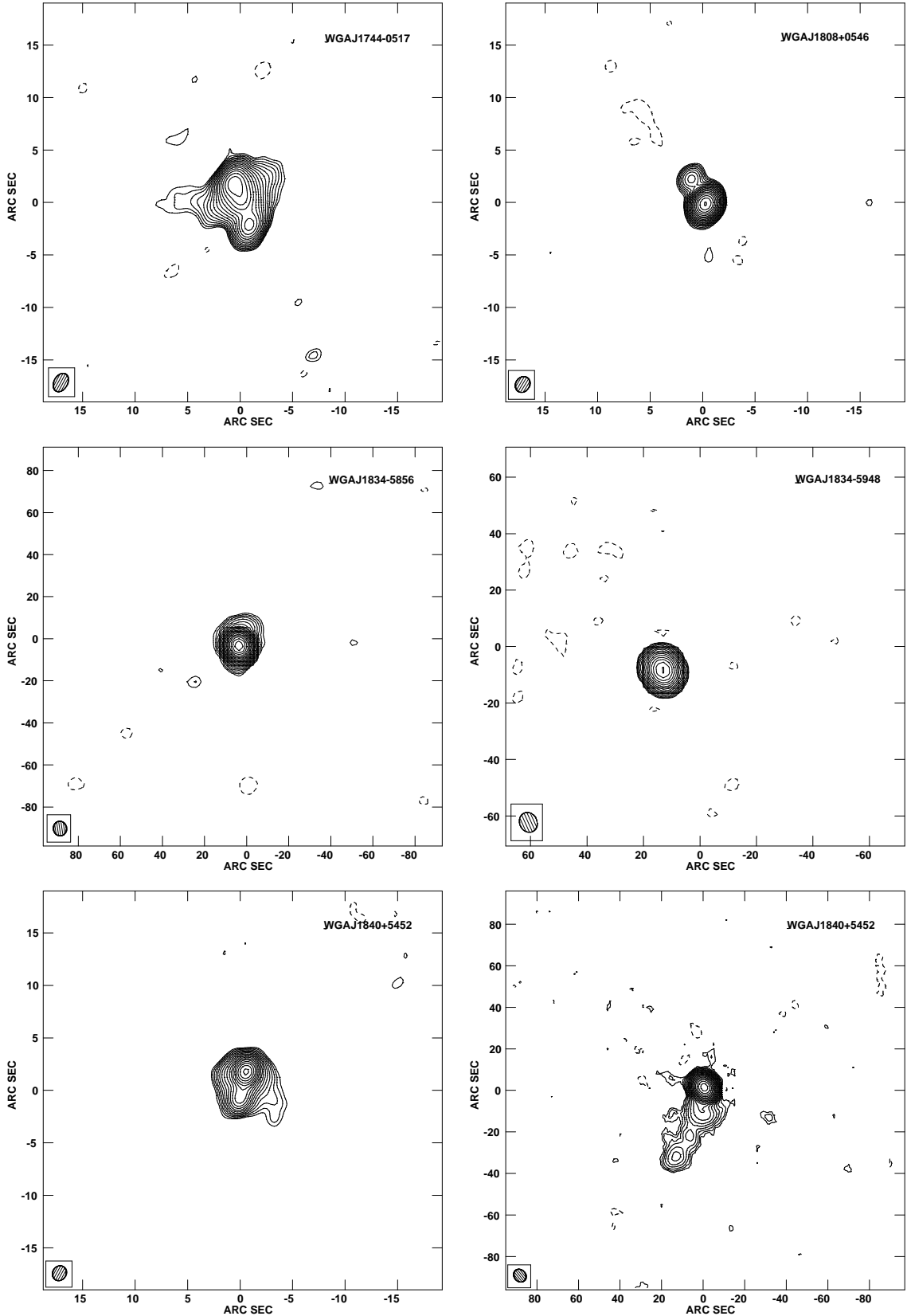

\centerline{
\includegraphics[scale=0.4]{A_WGAJ1744-0517.PS}
\includegraphics[scale=0.4]{A_WGAJ1808+0546.PS}
}
\centerline{
\includegraphics[scale=0.4]{6B_WGAJ1834-5856.PS}
\includegraphics[scale=0.4]{6B_WGAJ1834-5948.PS}
}
\centerline{
\includegraphics[scale=0.4]{A_WGAJ1840+5452.PS}
\includegraphics[scale=0.4]{AC_WGAJ1840+5452.PS}
}
\caption{\label{maps5} (a) WGA J1744$-$0517, VLA A. Image rms is 0.12
  mJy/beam. Image peak is 62.6 mJy/beam. (b) WGA J1808$+$0546, VLA
  A. Image rms is 0.19 mJy/beam. Image peak is 215.3 mJy/beam. (c) WGA
  J1834$-$5856, ATCA 6B. Image rms is 0.12 mJy/beam. Image peak is
  351.7 mJy/beam. (d) WGA J1834$-$5948, ATCA 6B. Image rms is 0.22
  mJy/beam. Image peak is 170.7 mJy/beam. (e) WGA J1840$+$5452, VLA
  A. Image rms is 0.12 mJy/beam. Image peak is 101.9 mJy/beam. (f) WGA
  J1840$+$5452, VLA A$+$C. Image rms is 0.11 mJy/beam. Image peak is
  139.9 mJy/beam. Contours and positive values as in
  Fig. \ref{maps1}.}
\end{figure*}


\begin{figure*}
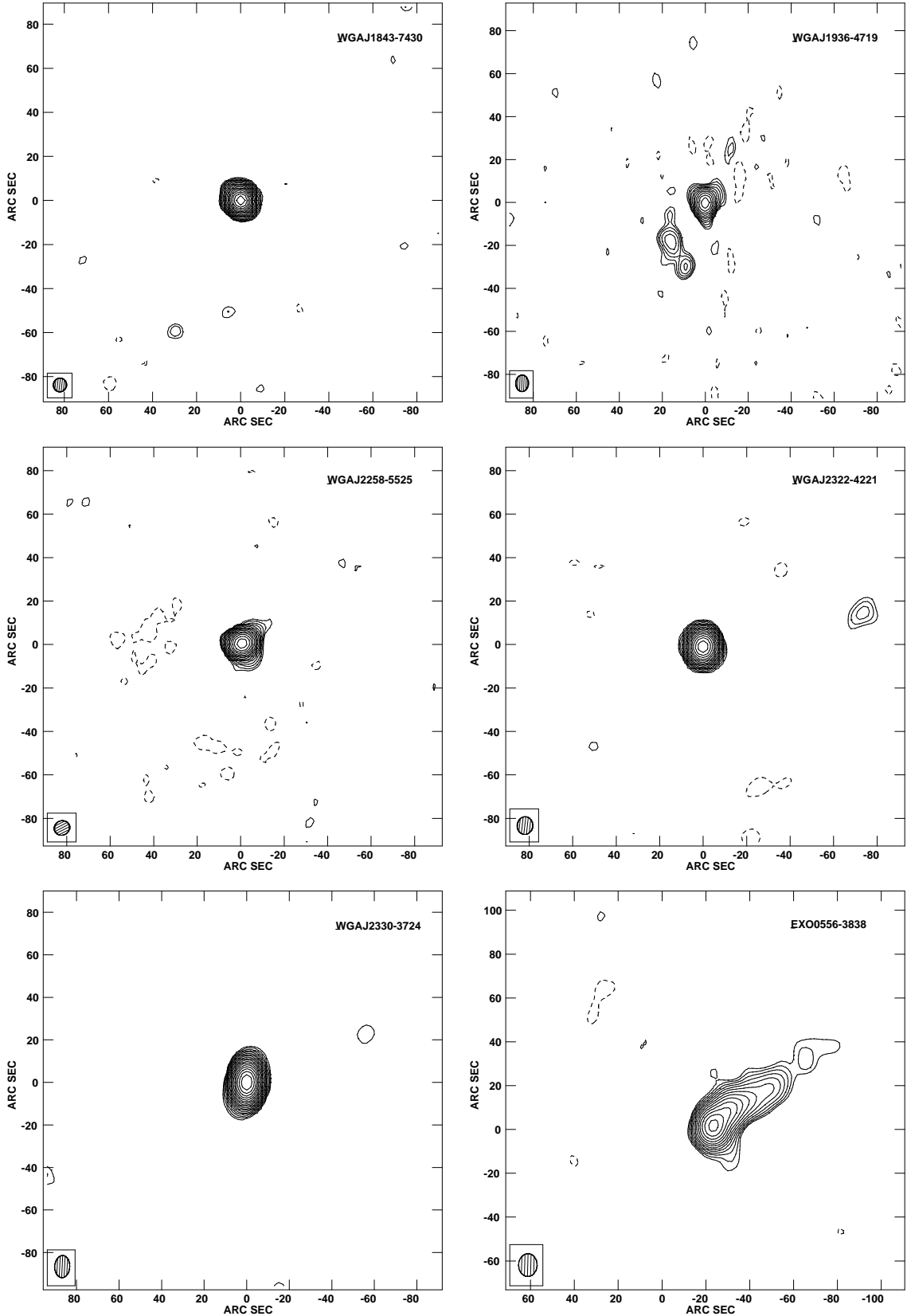

\centerline{
\includegraphics[scale=0.4]{6B_WGAJ1843-7430.PS}
\includegraphics[scale=0.4]{6B_WGAJ1936-4719.PS}
}
\centerline{
\includegraphics[scale=0.4]{6B_WGAJ2258-5525.PS}
\includegraphics[scale=0.4]{6B_WGAJ2322-4221.PS}
}
\centerline{
\includegraphics[scale=0.4]{6B_WGAJ2330-3724.PS}
\includegraphics[scale=0.4]{6A_EXO0556-3838.PS}
}
\caption{\label{maps6} (a) WGA J1843$-$7430, ATCA 6B. Image rms is
  0.07 mJy/beam. Image peak is 71.0 mJy/beam. (b) WGA J1936$-$4719,
  ATCA 6B. Image rms is 0.39 mJy/beam. Image peak is 98.7
  mJy/beam. (c) WGA J2258$-$5525, ATCA 6B. Image rms is 0.13
  mJy/beam. Image peak is 65.4 mJy/beam. (d) WGA J2322$-$4221, ATCA
  6B. Image rms is 0.10 mJy/beam. Image peak is 69.2 mJy/beam. (e) WGA
  J2330$-$3724, ATCA 6B. Image rms is 0.23 mJy/beam. Image peak is
  345.2 mJy/beam. (f) EXO 0556$-$3838, ATCA 6A. Image rms is 0.20
  mJy/beam. Image peak is 43.3 mJy/beam. Contours and positive values
  as in Fig. \ref{maps1}.}
\end{figure*}


\begin{figure*}
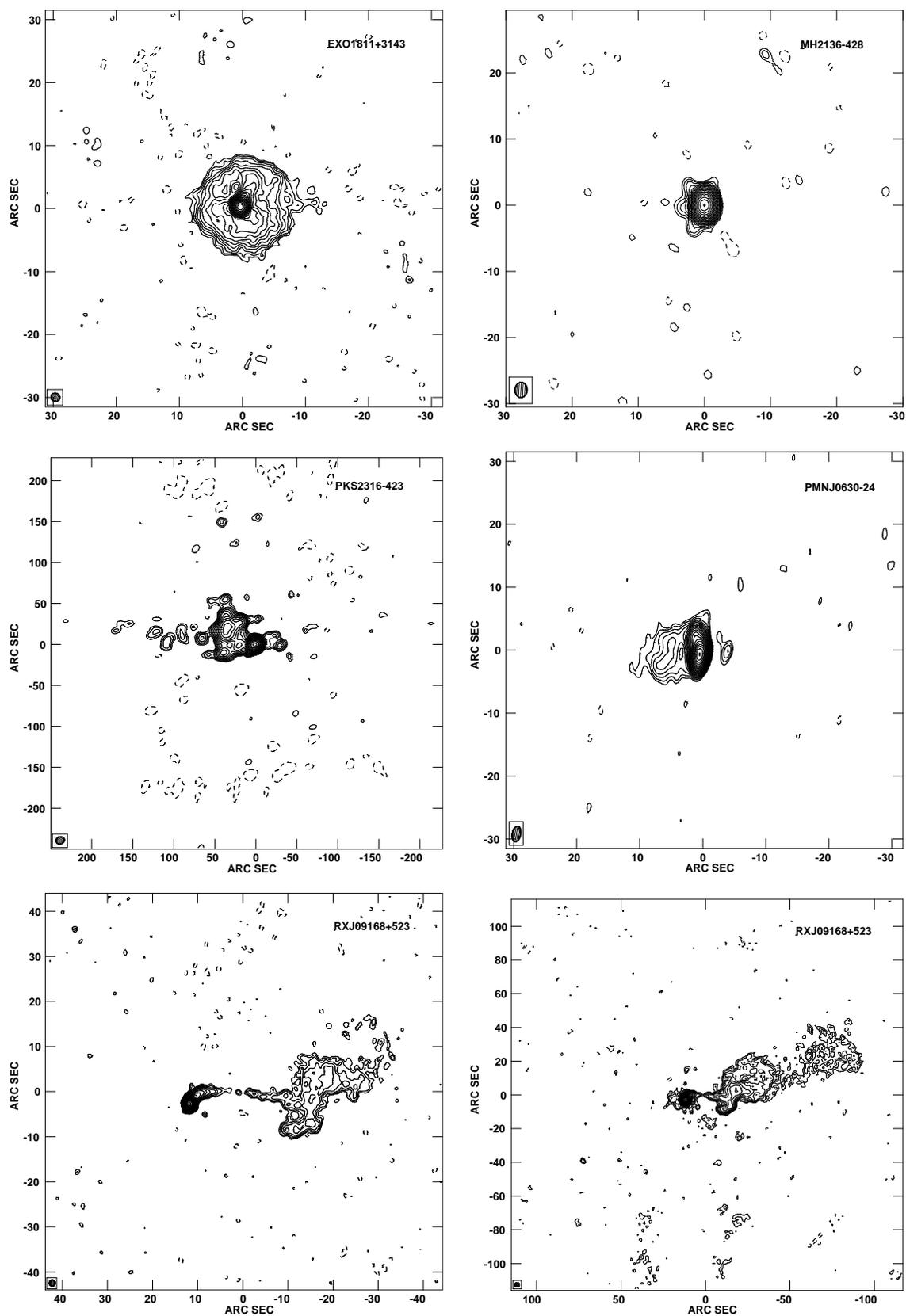

\centerline{
\includegraphics[scale=0.4]{A_EXO1811+3143.PS}
\includegraphics[scale=0.4]{6B_MH2136-428.PS}
}
\centerline{
\includegraphics[scale=0.4]{6B_PKS2316-423.PS}
\includegraphics[scale=0.4]{A_PMNJ0630-24.PS}
}
\centerline{
\includegraphics[scale=0.4]{A_RXJ09168+523.PS}
\includegraphics[scale=0.4]{AC_RXJ09168+523.PS}
}
\caption{\label{maps7} (a) EXO 1811$+$3143, VLA A. Image rms is 0.02
  mJy/beam. Image peak is 102.8 mJy/beam. (b) MH 2136$-$428, ATCA 6B
  at 4.8 GHz. Image rms is 0.04 mJy/beam. Image peak is 62.1
  mJy/beam. (c) PKS 2316$-$423, ATCA 6B. Image rms is 0.40
  mJy/beam. Image peak is 247.5 mJy/beam. (d) PMN J0630$-$24, VLA
  A. Image rms is 0.04 mJy/beam. Image peak is 87.1 mJy/beam. (e) RX
  J09168$+$523, VLA A. Image rms is 0.02 mJy/beam. Image peak is 57.8
  mJy/beam. (f) RX J09168$+$523, VLA A$+$C. Image rms is 0.04
  mJy/beam. Image peak is 61.1 mJy/beam. Contours and positive values
  as in Fig. \ref{maps1}.}
\end{figure*}


\begin{figure*}
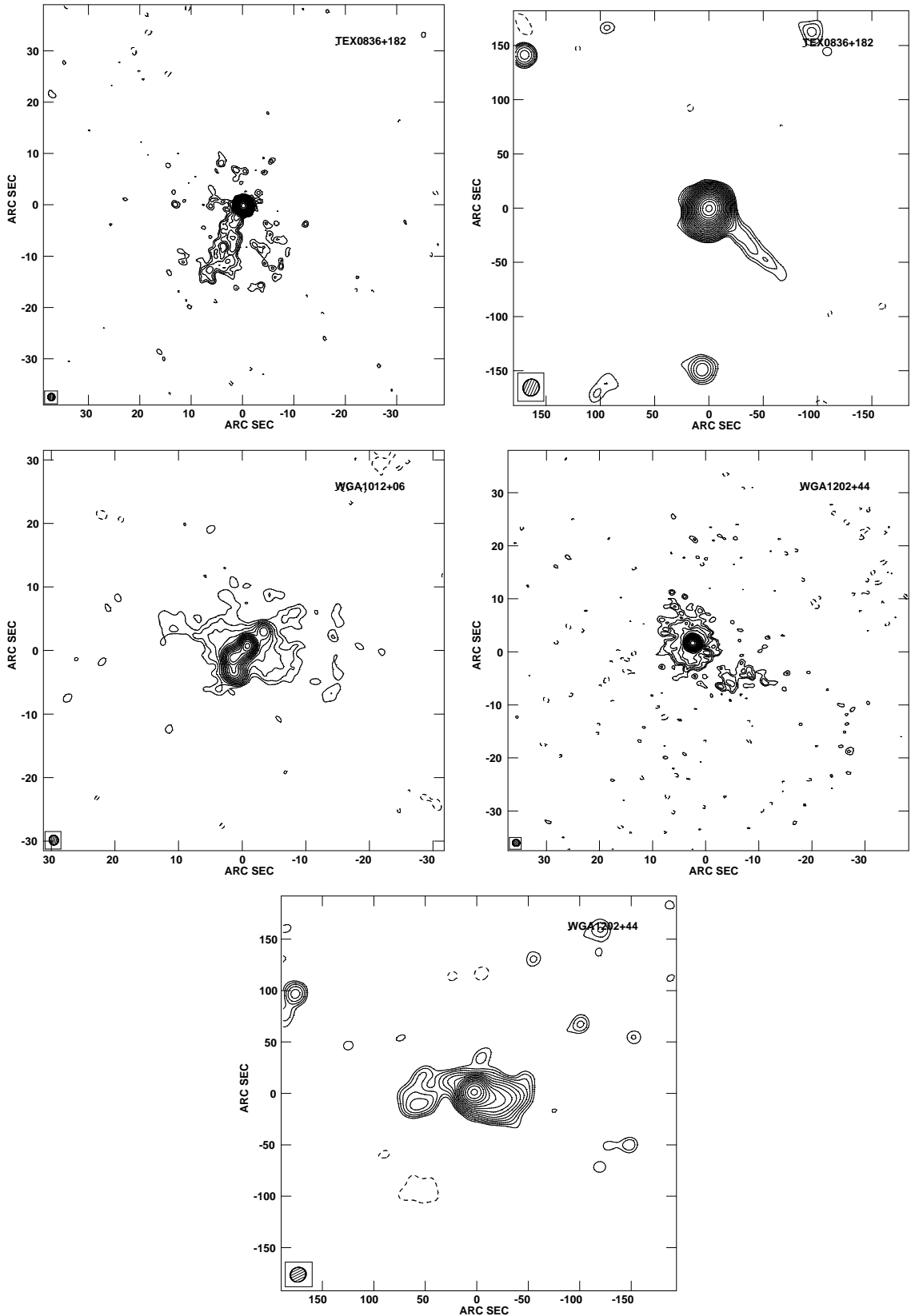

\centerline{
\includegraphics[scale=0.4]{A_TEX0836+182.PS}
\includegraphics[scale=0.4]{C_TEX0836+182.PS}
}
\centerline{
\includegraphics[scale=0.4]{A_WGA1012+06.PS}
\includegraphics[scale=0.4]{A_WGA1202+44.PS}
}
\centerline{
\includegraphics[scale=0.4]{C_WGA1202+44.PS}
}
\caption{\label{maps8} (a) TEX 0836$+$182, VLA A. Image rms is 0.08
  mJy/beam. Image peak is 308.3 mJy/beam. (b) TEX 0836$+$182, VLA
  C. Image rms is 0.11 mJy/beam. Image peak is 363.5 mJy/beam. (c) WGA
  1012$+$06, VLA A. Image rms is 0.36 mJy/beam. Image peak is 166.4
  mJy/beam. (d) WGA 1202$+$44, VLA A. Image rms is 0.03
  mJy/beam. Image peak is 48.5 mJy/beam. (e) WGA 1202$+$44, VLA
  C. Image rms is 0.14 mJy/beam. Image peak is 81.9 mJy/beam. Contours
  and positive values as in Fig. \ref{maps1}.}
\end{figure*}


\begin{table*}
\caption{\label{map} Map Parameters}
\begin{tabular}{lccrcrrc}
\hline
Object Name & Array & Beam & PA & rms & Peak & Scale & Figure \\
&& [arcsec] & [deg] & [mJy/ & [mJy/ & [kpc/ & \\
&&&& beam] & beam] & arcsec] & \\
(1) & (2) & (3) & (4) & (5) & (6) & (7) & (8) \\
\hline
WGA J0032$-$2849 & VLA A        &  3.0$\times$ 1.3 & $-$18 & 0.07 &  151.8 &  6.2 & 1a \\
WGA J0032$-$2849 & VLA A$+$C    & 11.3$\times$ 7.0 & $-$50 & 0.06 &  144.6 &  6.2 & 1b \\
WGA J0040$-$2340 & VLA C        & 15.2$\times$ 9.3 & $ $62 & 0.05 &   45.9 &  4.7 & 1c \\
WGA J0043$-$2638 & VLA A        &  3.3$\times$ 1.3 & $-$25 & 0.05 &   76.4 & 10.9 & 1d \\
WGA J0100$-$3337 & ATCA 6B      & 10.0$\times$ 7.1 & $-$ 1 & 0.20 &   90.1 & 10.4 & 1e \\
WGA J0245$+$1047 & VLA A        &  1.4$\times$ 1.4 & $ $75 & 0.04 &   28.6 &  1.8 & 1f \\
WGA J0245$+$1047 & VLA A$+$C    &  3.0$\times$ 2.6 & $-$66 & 0.04 &   30.5 &  1.8 & 2a \\
WGA J0313$+$4115 & VLA A        &  1.3$\times$ 1.2 & $ $28 & 0.17 &   54.8 &  0.8 & 2b \\
WGA J0428$-$3805 & ATCA 6A$+$6B & 10.7$\times$ 8.0 & $ $ 6 & 0.24 &  476.5 &  3.5 & 2c \\
WGA J0431$+$1731 & VLA A        &  1.5$\times$ 1.3 & $-$59 & 0.08 &  262.6 &  3.4 & 2d \\
WGA J0449$-$4349 & ATCA 6A$+$6B &  8.0$\times$ 6.5 & $ $ 3 & 0.18 &   99.3 &$>$4.0 & 2e \\
WGA J0528$-$5820 & ATCA 6A$+$6B &  8.9$\times$ 7.9 & $ $66 & 0.10 &   26.1 &  5.3 & 2f \\
WGA J0533$-$4632 & ATCA 6B      &  8.3$\times$ 6.1 & $ $26 & 0.11 &  108.2 &  6.3 & 3a \\
WGA J0558$+$5328 & VLA A        &  1.4$\times$ 1.3 & $ $19 & 0.07 &  296.9 &  1.0 & 3b \\
WGA J0558$+$5328 & VLA A$+$C    &  6.8$\times$ 5.1 & $ $33 & 0.06 &  355.1 &  1.0 & 3c \\
WGA J0624$-$3229 & VLA A        &  2.8$\times$ 1.4 & $ $ 3 & 0.06 &   34.8 &  5.2 & 3d \\
WGA J0847$+$1133 & VLA A        &  1.4$\times$ 1.3 & $ $28 & 0.05 &   26.9 &  4.4 & 3e \\
WGA J0940$+$2603 & VLA A        &  1.3$\times$ 1.2 & $ $15 & 0.06 &  371.7 &$>$8.6 & 3f \\
WGA J1204$-$0710 & VLA A        &  1.7$\times$ 1.4 & $ $ 5 & 0.08 &  148.1 &  4.1 & 4a \\
WGA J1231$+$2848 & VLA A        &  1.4$\times$ 1.3 & $-$ 3 & 0.02 &   85.6 &$>$10.4 & 4b \\
WGA J1231$+$2848 & VLA A$+$C    &  2.5$\times$ 2.1 & $-$38 & 0.04 &   90.5 &$>$10.4 & 4c \\
WGA J1311$-$0521 & VLA C        & 20.4$\times$15.7 & $ $38 & 0.07 &   79.1 &  3.7 & 4d \\
WGA J1320$+$0140 & VLA A        &  1.5$\times$ 1.5 & $-$ 5 & 0.07 &  496.5 & 11.6 & 4e \\
WGA J1539$-$0658 & VLA A        &  1.7$\times$ 1.4 & $-$23 & 0.02 &   43.5 &$>$8.3 & 4f \\
WGA J1744$-$0517 & VLA A        &  1.9$\times$ 1.3 & $-$28 & 0.12 &   62.6 &  6.1 & 5a \\
WGA J1808$+$0546 & VLA A        &  1.6$\times$ 1.3 & $-$31 & 0.19 &  215.3 &  4.1 & 5b \\
WGA J1834$-$5856 & ATCA 6B      &  7.0$\times$ 6.2 & $ $ 8 & 0.12 &  351.7 &$>$8.8 & 5c \\
WGA J1834$-$5948 & ATCA 6B      &  7.2$\times$ 6.2 & $ $24 & 0.22 &  170.7 &  7.5 & 5d \\
WGA J1840$+$5452 & VLA A        &  1.5$\times$ 1.3 & $-$29 & 0.12 &  101.9 &  9.2 & 5e \\
WGA J1840$+$5452 & VLA A$+$C    &  6.6$\times$ 5.5 & $ $38 & 0.11 &  139.9 &  9.2 & 5f \\
WGA J1843$-$7430 & ATCA 6B      &  6.3$\times$ 5.9 & $-$ 6 & 0.07 &   71.0 &  3.8 & 6a \\
WGA J1936$-$4719 & ATCA 6B      &  7.7$\times$ 5.8 & $-$ 2 & 0.39 &   98.7 &  5.4 & 6b \\
WGA J2258$-$5525 & ATCA 6B      &  7.2$\times$ 6.6 & $-$63 & 0.13 &   65.4 &  7.9 & 6c \\
WGA J2322$-$4221 & ATCA 6B      &  8.2$\times$ 7.1 & $-$ 7 & 0.10 &   69.2 &  2.3 & 6d \\
WGA J2330$-$3724 & ATCA 6B      & 10.4$\times$ 7.0 & $-$ 3 & 0.23 &  345.2 &  5.6 & 6e \\
EXO 0556$-$3838  & ATCA 6A      & 10.4$\times$ 8.4 & $-$ 2 & 0.20 &   43.3 &  6.0 & 6f \\
EXO 1811$+$3143  & VLA A        &  1.4$\times$ 1.3 & $ $42 & 0.02 &  102.8 &  2.9 & 7a \\
MH 2136$-$428$^\star$& ATCA 6B  &  2.3$\times$ 1.8 & $-$ 3 & 0.04 &   62.1 &$>$5.4 & 7b \\
PKS 2316$-$423   & ATCA 6B      & 10.1$\times$ 8.8 & $-$60 & 0.40 &  247.5 &  1.5 & 7c \\
PMN J0630$-$24   & VLA A        &  2.5$\times$ 1.3 & $-$10 & 0.04 &   87.1 & 11.6 & 7d \\
RX J09168$+$523  & VLA A        &  1.4$\times$ 1.2 & $ $ 2 & 0.02 &   57.8 &  4.3 & 7e \\
RX J09168$+$523  & VLA A$+$C    &  2.6$\times$ 2.3 & $ $14 & 0.04 &   61.1 &  4.3 & 7f \\
TEX 0836$+$182   & VLA A        &  1.4$\times$ 1.3 & $ $ 6 & 0.08 &  308.3 &$>$7.8 & 8a \\
TEX 0836$+$182   & VLA C        & 15.8$\times$14.5 & $-$24 & 0.11 &  363.5 &$>$7.8 & 8b \\
WGA 1012$+$06    & VLA A        &  1.5$\times$ 1.4 & $ $13 & 0.36 &  166.4 &  9.7 & 8c \\
WGA 1202$+$44    & VLA A        &  1.3$\times$ 1.3 & $ $52 & 0.03 &   48.5 &  5.9 & 8d \\
WGA 1202$+$44    & VLA C        & 15.5$\times$14.0 & $-$63 & 0.14 &   81.9 &  5.9 & 8e \\
\hline
\end{tabular}

\medskip

\parbox[]{11.5cm}{The columns are: (1) object name; (2) telescope
configuration; (3) beam size; (4) position angle; (5) image rms; (6)
peak flux; (7) physical scale in the object's rest-frame; and (8)
corresponding panel in Figs. \ref{maps1}-\ref{maps8}.}

\medskip

\parbox[]{11.5cm}{$^\star$ observed at 4.8 GHz}

\end{table*}


We have observed the northern ($\delta \ga -30^{\circ}$) sources in
the sample (24 objects) with the VLA and the southern sources (15
objects) with the ATCA. In Table \ref{general} we give the log of our
observations. Another five northern sources have published radio data
useful for our purpose and were not reobserved. These are listed in
Table \ref{generalpub} with the corresponding references.

The VLA observations were conducted in continuum mode whereby two
channels of 50 MHz bandwidth were centered at 1.385 and 1.465 GHz,
resulting in an effective frequency of 1.425 GHz. During a period of
several years, between January 2002 and May 2006, we imaged all
sources with both the VLA A and C configurations. The only exception
was the source WGA J1808$+$0546, which we imaged only with the VLA A
array. Typically three or four scans of several minutes length spaced
to optimize the $(u,v)$ plane coverage were interleaved with one
minute scans on a suitable secondary VLA calibrator for each
source. This yielded total source exposure times of $\sim 30 - 45$
minutes, corresponding to a theoretical rms sensitivity of $\sim 0.05$
mJy/beam. Multiple observations of 3C 48, 3C 147, or 3C 286 were used
to flux-calibrate the maps.

The VLA data were processed with the Astronomical Image Processing
System (AIPS; version 31DEC06) package. A model based on Clean
components was used to start the self-calibration process. Phase-only
self-calibration was used for the first three iterations and amplitude
and phase self-calibration for the last one or two iterations. The
task IMAGR with robust weighting (\mbox{ROBUST}=0.5) was used to
generate the maps and Clean components. We found it useful to combine
the self-calibrated data sets from the A and C arrays in order to
increase the sensitivity and to improve the $(u,v)$ plane coverage in
only 6/23 cases.

The ATCA continuum observations were performed with a bandwidth of 128
MHz centered at 1.384 GHz. The observations were carried out in two 13
hour sessions in May 2002 in the 6A configuration and in a single
session of 32 hours in January 2004 in the 6B configuration. Typically
sources were observed in scans of $\sim 10 - 15$ minutes spread over a
wide range of hour angles, interleaved with two minute scans on a
nearby secondary calibrator. The total integration time on each source
ranged between $\sim 2 - 5$ hours, resulting in theoretical rms
sensitivity of $\sim 0.05$ mJy/beam. The rms noise in the final images
was a factor of two or more larger than this due to the presence of
confusing sources in the primary beam. The source MH 2136$-$428 was
included in a different program and was observed instead at 4.8
GHz. The ATCA data reduction was performed using standard procedures
in the MIRIAD software package. The images were produced in either
MIRIAD or Difmap using a similar method to that used for the VLA
data. The flux density scale is tied to the ATCA primary calibrator
PKS 1934$-$638.

The final maps are shown in Figs. \ref{maps1} - \ref{maps8} and the
corresponding map parameters are listed in Table \ref{map}. The
dynamic ranges (peak/noise) of the VLA maps lie between $\sim
300-7000$ and those of the ATCA maps between $\sim 200-3000$, with
typical values of $\sim 2000$ and 1000, respectively. In Table
\ref{radioprop} we list the radio properties of the sources as derived
from our observations using the AIPS task \mbox{TVSTAT}. The core flux
density was measured as the peak of the core emission on the VLA A
array map and (combined) ATCA map for the northern and southern
sources, respectively. The extended flux density was derived by
subtracting the core flux density from the total flux density, the
latter measured on the VLA C array map, if the source was resolved,
else measured on the VLA A array map, for the northern sources, and on
the (combined) ATCA map for the southern sources. We give $1\sigma$
cumulative errors for the extended flux densities, which depend upon
the solid angular extent of the measured flux.

We have detected extended radio emission for all sources observed with
the VLA, with the exception of WGA J0043$-$2638. For about half of
these objects (13/24 sources) the extended emission is visible only on
the small scales typical of the A array and the source remains
unresolved on the C array map. Similarly, 6/15 sources (or 40 per
cent) observed with the ATCA remain unresolved on the scales
imaged. Since the angular resolution of the ATCA is intermediate
between that of the VLA A and C array, the non-detection of extended
radio emission in these sources is most likely due to resolution
rather than to sensitivity. Therefore, we have derived upper limits on
the extended radio emission for the seven unresolved sources in our
sample by assuming the most basic definition of a blazar, i.e., a
radio core-dominance parameter $R=L_{\rm core}/L_{\rm ext}>1$, where
$L_{\rm core}$ and $L_{\rm ext}$ are the radio core and extended
luminosities, respectively. For these sources the core flux densities
listed in Table \ref{radioprop}, column (5), are then assumed to be
the {\it total} flux densities.


\begin{table*}
\caption{\label{radioprop} 
Observed 1.4 GHz Radio Properties of the Sample}
\begin{tabular}{lcllrrrrrrc}
\hline
Object Name & Array & R.A.(J2000) & Decl.(J2000) & $f_{\rm core}$ & $f_{\rm ext}$ & 
$\log L_{\rm core}$ & $\log L_{\rm ext}$ & $\log R$ & LAS & log \\
&&&& [mJy] & [mJy] & [W/Hz] & [W/Hz] && [arcsec] & $(L_{\rm core}/L_{\rm x})$ \\
(1) & (2) & (3) & (4) & (5) & (6) & (7) & (8) & (9) & (10) & (11) \\
\hline
WGA J0032$-$2849 & VLA  & 00 32 33.081 & $-$28 49 20.00 & 151.8 &  12.2$\pm$0.3 & 25.84 & 24.85 & $ $0.99 &  34.6 & 6.36 \\
WGA J0040$-$2340 & VLA  & 00 40 24.892 & $-$23 40 00.80 &  45.7 &   3.7$\pm$0.2 & 24.95 & 23.93 & $ $1.02 &  18.4 & 6.60 \\
WGA J0043$-$2638 & VLA  & 00 43 22.731 & $-$26 39 06.10 &  76.4 &           ... & 26.57 &$<$26.27&$>$0.00 &$<$3.3 & 5.73 \\
WGA J0100$-$3337 & ATCA & 01 00 09.400 & $-$33 37 31.00 &  90.1 &           ... & 26.51 &$<$26.21&$>$0.00 &$<$10.0& 6.62 \\
WGA J0245$+$1047 & VLA  & 02 45 13.473 & $+$10 47 22.70 &  28.6 & 408.7$\pm$0.9 & 23.78 & 24.96 & $-$1.18 & 104.1 & 5.00 \\
WGA J0313$+$4115 & VLA  & 03 13 57.649 & $+$41 15 23.85 &  54.8 & 133.0$\pm$1.9 & 23.30 & 23.69 & $-$0.39 &  14.2 & 5.68 \\
WGA J0428$-$3805 & ATCA & 04 28 50.800 & $-$38 05 50.00 &  29.7 &  32.1$\pm$2.1 & 24.46 & 24.54 & $-$0.08 &  68.1 & 6.80 \\
WGA J0431$+$1731 & VLA  & 04 31 57.367 & $+$17 31 35.90 & 262.6 & 113.1$\pm$0.5 & 25.37 & 25.05 & $ $0.32 &   2.8 & 6.81 \\
WGA J0449$-$4349 & ATCA & 04 49 24.700 & $-$43 50 08.00 &  99.3 & 183.2$\pm$1.9 &$>$25.12&$>$25.45&$-$0.33&  16.8 & 5.23 \\
WGA J0528$-$5820 & ATCA & 05 28 34.699 & $-$58 20 17.99 &  26.1 & 128.2$\pm$1.3 & 24.87 & 25.64 & $-$0.77 &  54.2 & 5.81 \\
WGA J0533$-$4632 & ATCA & 05 33 40.799 & $-$46 32 15.00 & 108.2 &           ... & 25.72 &$<$25.42&$>$0.00 &$<$8.3 & 6.54 \\
WGA J0558$+$5328 & VLA  & 05 58 11.822 & $+$53 28 17.75 & 296.9 &  88.0$\pm$0.3 & 24.22 & 23.70 & $ $0.52 &  31.5 & 6.85 \\
WGA J0624$-$3229 & VLA  & 06 24 45.058 & $-$32 30 54.85 &  34.8 &  11.4$\pm$0.4 & 24.97 & 24.57 & $ $0.40 &  11.8 & 5.10 \\
WGA J0847$+$1133 & VLA  & 08 47 12.940 & $+$11 33 50.25 &  26.9 &  10.2$\pm$0.2 & 24.66 & 24.30 & $ $0.36 &   1.5 & 3.74 \\
WGA J0940$+$2603 & VLA  & 09 40 14.720 & $+$26 03 30.00 & 371.7 &  97.3$\pm$0.5 &$>$26.73&$>$26.31&$ $0.42&   6.0 & 6.42 \\
WGA J1204$-$0710 & VLA  & 12 04 16.656 & $-$07 10 09.25 & 148.1 &  15.5$\pm$0.3 & 25.33 & 24.41 & $ $0.92 &   3.0 & 6.24 \\
WGA J1231$+$2848 & VLA  & 12 31 43.565 & $+$28 47 49.70 &  85.6 &  61.6$\pm$0.7 &$>$26.50&$>$26.57&$-$0.07&  24.0 & 5.43 \\
WGA J1311$-$0521 & VLA  & 13 11 17.840 & $-$05 21 20.00 &  27.7 &  42.2$\pm$0.4 & 24.49 & 24.72 & $-$0.23 &  94.8 & 5.77 \\
WGA J1320$+$0140 & VLA  & 13 20 26.787 & $+$01 40 36.75 & 496.5 & 198.6$\pm$0.3 & 27.58 & 27.46 & $ $0.12 &   3.6 & 6.72 \\
WGA J1539$-$0658 & VLA  & 15 39 09.677 & $-$06 58 43.15 &  43.5 &   2.2$\pm$0.2 &$>$25.73&$>$24.58&$ $1.15&   6.0 & 6.16 \\
WGA J1744$-$0517 & VLA  & 17 44 20.883 & $-$05 18 39.55 &  62.6 & 135.8$\pm$0.7 & 25.42 & 25.85 & $-$0.43 &   4.2 & 6.27 \\
WGA J1808$+$0546 & VLA  & 18 08 32.223 & $+$05 46 51.85 & 215.3 &  16.4$\pm$0.7 & 25.48 & 24.42 & $ $1.06 &   2.8 & 7.10 \\
WGA J1834$-$5856 & ATCA & 18 34 27.479 & $-$58 56 36.71 & 351.7 &  16.0$\pm$0.6 &$>$26.75&$>$25.57&$ $1.18 &  9.3 & 6.29 \\
WGA J1834$-$5948 & ATCA & 18 34 15.324 & $-$59 48 46.43 & 170.7 &           ... & 26.16 &$<$25.86&$>$0.00 &$<$7.2 & 6.25 \\
WGA J1840$+$5452 & VLA  & 18 40 57.382 & $+$54 52 15.85 & 101.9 &  75.2$\pm$0.3 & 26.29 & 26.33 & $-$0.04 &  29.7 & 5.50 \\
WGA J1843$-$7430 & ATCA & 18 43 40.198 & $-$74 30 24.99 &  71.0 &           ... & 24.93 &$<$24.63&$>$0.00 &$<$6.3 & 7.07 \\
WGA J1936$-$4719 & ATCA & 19 36 56.100 & $-$47 19 50.00 &  49.2 &  25.4$\pm$2.1 & 25.17 & 24.97 & $ $0.20 &  30.5 & 4.59 \\
WGA J2258$-$5525 & ATCA & 22 58 19.098 & $-$55 25 37.99 &  65.4 &   7.9$\pm$0.5 & 25.83 & 25.04 & $ $0.79 &   9.5 & 5.31 \\
WGA J2322$-$4221 & ATCA & 23 22 40.398 & $-$42 20 43.89 &  69.2 &           ... & 24.37 &$<$24.07&$>$0.00 &$<$8.2 & 6.51 \\
WGA J2330$-$3724 & ATCA & 23 30 35.798 & $-$37 24 37.00 & 345.2 &           ... & 26.07 &$<$25.77&$>$0.00 &$<$10.4& 6.83 \\
B2 1147$+$245    & VLA  & 11 50 19.22  & $+$24 17 53.7  & 664   &  25           &$>$26.39&$>$25.06&$ $1.33&  26.6 & 7.34 \\
EXO 0556$-$3838  & ATCA & 05 58 06.477 & $-$38 38 31.12 &  43.3 &  43.0$\pm$1.1 & 25.24 & 25.33 & $-$0.09 &  22.9 & 4.02 \\
EXO 1811$+$3143  & VLA  & 18 13 35.190 & $+$31 44 17.65 & 102.8 &  64.0$\pm$0.3 & 24.78 & 24.62 & $ $0.16 &   3.8 & 6.00 \\
MH 2136$-$428$^\star$&ATCA&21 39 24.140&$-$42 35 21.30&  62.1 &   1.4$\pm$0.2 &$>$25.27&$>$23.70& $ $1.57 &   2.5 & 5.62 \\
ON 325           & VLA  & 12 17 52.11  & $+$30 07 00.0  & 355   & 189           & 25.94 & 25.74 & $ $0.20 &  95.0 & 5.77 \\
PKS 2316$-$423   & ATCA & 23 19 05.963 & $-$42 06 49.00 & 247.5 & 391.9$\pm$4.9 & 24.51 & 24.73 & $-$0.22 & 169.8 & 5.84 \\
PMN J0630$-$24   & VLA  & 06 30 59.527 & $-$24 06 46.05 &  87.1 &  16.9$\pm$0.3 & 26.83 & 26.40 & $ $0.43 &   6.5 & 5.01 \\
RX J09168$+$523  & VLA  & 09 16 51.921 & $+$52 38 28.45 &  57.8 &  77.6$\pm$0.4 & 24.96 & 25.14 & $-$0.18 &  95.6 & 5.12 \\
TEX 0836$+$182   & VLA  & 08 39 30.712 & $+$18 02 47.05 & 308.3 &  88.3$\pm$0.5 &$>$26.47&$>$26.06&$ $0.41&  90.3 & 6.38 \\
WGA 1012$+$06    & VLA  & 10 12 13.360 & $+$06 30 57.05 & 166.4 & 424.1$\pm$3.6 & 26.61 & 27.21 & $-$0.60 &   8.8 & 6.20 \\
WGA 1202$+$44    & VLA  & 12 02 08.665 & $+$44 44 22.55 &  48.5 &  67.4$\pm$1.0 & 25.27 & 25.51 & $-$0.24 &  57.2 & 5.50 \\
1ES 1212$+$078   & VLA  & 12 15 10.88  & $+$07 32 05.2  &  85   &  65           & 24.79 & 24.72 & $ $0.07 &       & 4.89 \\
3C 66A           & VLA  & 02 22 39.48  & $+$43 02 08.4  & 814   &1052           & 26.85 & 27.09 & $-$0.24 &  43.4 & 5.60 \\
4C $+$55.17      & VLA  & 09 57 38.18  & $+$55 22 57.4  &2568   & 381           & 28.00 & 27.39 & $ $0.60 &   3.1 & 7.55 \\
\hline
\end{tabular}

\medskip

\parbox[]{16.5cm}{The columns are: (1) object name; (2) telescope used
for observations; (3) and (4) position of the unresolved core,
measured on VLA A array or (combined) ATCA map (NVSS position is given
for sources listed in Table \ref{generalpub}); (5) core flux density,
measured on VLA A array or (combined) ATCA map; (6) extended flux
density, derived from the total flux density measured on VLA C array
map (if the source was resolved, else measured on VLA A array map) or
(combined) ATCA map substracting the core flux density given in column
(5); (7) core luminosity $k$-corrected with spectral index
$\alpha_{\rm r}=0$; (8) extended luminosity $k$-corrected with
spectral index $\alpha_{\rm r}=0.8$; (9) radio core-dominance
parameter $R$, defined as $R=L_{\rm core}/L_{\rm ext}$, where $L_{\rm
core}$ and $L_{\rm ext}$ are the core and extended luminosities,
respectively; (10) largest angular size (LAS), measured from the core
to the peak in the extended radio structure [following \citet{Mur93}]
on VLA C array map (if the source was resolved, else measured on VLA A
array map) or (combined) ATCA map; and (11) ratio of core luminosity
to total X-ray luminosity at 1 keV, the latter $k$-corrected with
spectral index $\alpha_{\rm x}=1.2$.}

\medskip

\parbox[]{16.5cm}{$^\star$ observed 4.8 GHz radio properties}

\end{table*}

\section{The Parent Population}

Current unified schemes for radio-loud AGN posit that BL Lacs are the
relativistically beamed counterparts of the low-luminosity FR I radio
galaxies, i.e., that they are FR Is with their radio jets aligned
close to our line of sight \citep[][and references
therein]{Urry95}. Since at small viewing angles orientation and
projection effects make the recognition of the FR class of the
extended radio morphology difficult, this unification scheme relies
mainly on comparisons between the extended radio powers of FR Is and
BL Lacs, which, believed to be isotropic and therefore unaffected by
relativistic beaming, are expected to be similar.

Nevertheless, as high-quality radio imaging campaigns have shown, the
connection between BL Lacs and radio galaxies seems to be more
complicated. BL Lacs with high, FR II-like extended radio powers
appear to be present in considerable numbers in surveys with
relatively high radio flux limits, such as, e.g., the 1-Jy survey
\citep{Kol92, Mur93, Cas99, Rec01}. In this section we show that also
in the DXRBS, a survey with a roughly ten times lower radio flux limit
than the 1-Jy survey, numerous BL Lacs can be associated with
relativistically beamed FR IIs (Sections \ref{lext} and \ref{morph}).
As we discuss in Section \ref{incon}, the existence of FR II BL Lacs
is in fact {\it expected} considering the current classification
scheme for radio-loud AGN and calls for a revision of the unified
schemes.

\subsection{The Distribution of Extended Radio Powers} \label{lext}


\begin{figure}
\centerline{
\includegraphics[scale=0.43]{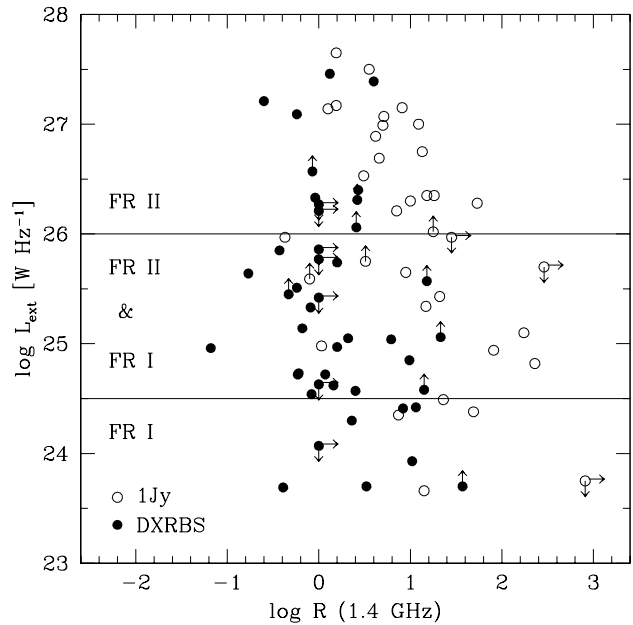}
}
\caption{\label{logRlext} The extended radio power versus the radio
core dominance parameter $R$ at 1.4 GHz, where $R=L_{\rm core}/L_{\rm
ext}$, with $L_{\rm core}$ and $L_{\rm ext}$ the radio core and
extended luminosities, respectively. Open and filled circles indicate
BL Lacs from the 1Jy survey and the DXRBS, respectively. Arrows
indicate limits. The horizontal solid lines mark the regions of
extended radio powers typical of FR I and FR II radio galaxies.}
\end{figure}

In Fig. \ref{logRlext} we have plotted for the DXRBS BL Lac sample
(filled circles) the radio core-dominance parameter $R$ versus the
extended radio luminosity at 1.4 GHz. The figure shows that whereas
the extended radio powers of DXRBS BL Lacs span a relatively wide
range ($\sim 4$ orders of magnitude), a considerable number of sources
(11/44 objects or 25 per cent) have values typical of FR II radio
galaxies \citep[$L_{\rm ext} > 10^{26}$ W Hz$^{-1}$;][]{Owen94} and
more than half of the sample (25/44 objects or 57 per cent) lies in
the regime occupied by both FR Is and FR IIs \citep[$10^{24.5} <
L_{\rm ext} < 10^{26}$ W Hz$^{-1}$;][]{Owen94}. This means that {\it a
large fraction of the sample are potentially beamed FR IIs}. Regarding
only the complete sample the numbers are similar, with 9/31 objects
(or 29 per cent) and 18/31 objects (or 58 per cent) having FR II and
FR I/II extended radio powers, respectively.

For two and four of the unresolved sources in our sample the
calculated upper limits on the extended radio emission fall in the FR
II and FR I/II regime, respectively. If these sources have values of
the radio core-dominance parameter $R\gg1$, the derived limits might
be highly overestimated. In this respect we note that if we use
instead the maximum radio core-dominance value observed for the entire
sample ($R\sim40$), the extended radio emission of three of these
sources reduces to values typical of FR Is.

In Fig. \ref{logRlext} we have included for comparison also the
\mbox{1-Jy} BL Lac sample \citep{Sti91, Rec01} (open circles), which
to date is the only other sizeable, complete radio flux-limited sample
besides the DXRBS. Radio core and extended fluxes at 1.4 GHz are
available from high-quality VLA observations for all 37 1-Jy BL Lacs
\citep[e.g.,][]{Mur93, Cas99, Rec01}, with the exception of the most
southern source (1Jy~2005$-$489). Three sources, namely,
1Jy~0048$-$097, 1Jy~0716$+$714, and 1Jy~2150$+$173, have featureless
optical spectra \citep{Rec01}, and we have used lower limits on the
redshifts (see Section \ref{red}) to calculate luminosities. We note
that one source (B2~1147$+$245) is common to both the 1-Jy and DXRBS
BL Lac samples. Fig. \ref{logRlext} shows that the majority in the
1-Jy sample is shifted towards higher extended radio powers relative
to the DXRBS. Roughly half the sample (18/37 objects or 49 per cent)
has values typical of FR II radio galaxies and more than a third
(13/37 objects or 35 per cent) lies in the regime occupied by both FR
Is and FR IIs. This shift can be attributed mainly to the different
radio flux limits and sky coverages of the two surveys. Both these
values are significantly higher for the 1-Jy survey than for the DXRBS
(radio flux limits at 5 GHz of 1 Jy and $\sim 50$ mJy, and sky
coverages of $\sim 10$ and $<1$ sr, respectively), thus favoring the
selection of high-power sources.

\subsection{The Radio Morphologies} \label{morph}

In the previous subsection we have shown that a large majority ($\sim
80$ per cent) of the DXRBS BL Lac sample have extended radio powers
typical of FR II radio galaxies. Is also their radio morphology
consistent with them being relativistically beamed FR IIs? An answer
to this question is especially important for that part of the sample
with extended radio powers in the overlap regime between FR Is and FR
IIs.

The main defining feature of an FR II radio galaxy is the presence of
well-collimated jets that terminate in 'hot spots', i.e., small areas
of high surface brightness. By definition, these hot spots are located
towards the ends of the extended radio lobes \citep{Fan74}. However,
the radio morphology expected for an FR II viewed at small angles is
not entirely clear. The emission from the lobes may overlap in
projection giving the appearance of an extended halo. On the near-side
a relatively well-collimated, one-sided jet terminating in a hot spot
should appear. In some cases we might also expect to see an isolated
hot spot on the other side of the amplified jet, due to the (presumed)
counterjet. These features are generally observed in steep- and
flat-spectrum radio quasars \citep[e.g.,][]{Bro82, Kol90, Mur93,
Kap98, L06}, which are believed to be the relativistically beamed
counterparts of FR II radio galaxies. Do we see some or all of them
also in the DXRBS BL Lacs?

\subsubsection{BL Lacs with FR II Extended Radio Powers}

Eleven DXRBS BL Lacs have FR II-like extended radio powers ($L_{\rm
ext} > 10^{26}$ W Hz$^{-1}$). Two of these sources, namely, WGA
J0043$-$2638 (Fig. 1d) and WGA J0100$-$3337 (Fig. 1e), are unresolved
on the scales imaged and their association with an FR class remains
uncertain. For the remaining nine sources we discuss in the following
their radio morphology individually and in decreasing order of the
extended radio luminosity.

\vspace*{0.1cm}

{\it WGA J1320$+$0140. $-$} The VLA A array map (Fig. 4e) shows the
core region to be elongated to the west. Most of the diffuse radio
emission is observed perpendicular to the axis of the core elongation
and extends to one side only. It has a broad jet-like structure, which
appears to be embedded in a lobe. Some low-level extended radio
emission is found also along the axis of the core elongation.

\vspace*{0.1cm}

{\it 4C $+$55.17. $-$} The VLA A array map of \citet{Mur93} shows a
one-sided jet-like structure and a diffuse extended radio halo to both
sides of the core-jet axis. A 408 MHz radio map of similar resolution
taken by \citet{Bro82} shows the one-sided jet to be relatively
well-collimated and a potential hot spot can be seen on the other side
of the core.

\vspace*{0.1cm}

{\it WGA 1012$+$06. $-$} The VLA A array map (Fig. 8c) shows a
one-sided jet emanating from the core, which first bends to the
south-east and then to the south. A possible hot spot of a counterjet
is seen on the opposite side of the core. The core-jet structure
appears to be embedded in a diffuse extended radio halo.

\vspace*{0.1cm}

{\it 3C 66A. $-$} The VLA BnA array maps of \citet{Ulv83} show a short
(a few arcseconds long) one-sided jet. The core-jet structure is
surrounded by a large-scale diffuse radio halo.

\vspace*{0.1cm}

{\it WGA J1231$+$2848. $-$} The VLA A array map (Fig. 4b) shows a
short ($\sim 5''$ long) one-sided jet emerging from the core and
ending in a hot spot. The combined VLA A$+$C array map (Fig. 4c) shows
that this core-jet structure is embedded in a large-scale diffuse lobe
emission. A low-level counterjet seems to be present on the opposite
side of the core.

\vspace*{0.1cm}

{\it PMN J0630$-$24. $-$} The VLA A array map (Fig. 7d) shows the core
region to be elongated to the north. Most of the diffuse radio
emission is observed perpendicular to the axis of the core elongation
and extends to one side only. It has a broad jet-like structure. Some
low-level extended radio emission is found to the opposite side of the
large-scale extended radio structure.

\vspace*{0.1cm}

{\it WGA J1840$+$5452. $-$} The VLA A array map (Fig. 5e) shows a
short jet-like structure emerging from the core. The jet is observed
on the VLA A$+$C array map (Fig. 5f) to continue over scales of tens
of arcseconds, possibly ending in a hot spot.

\vspace*{0.1cm}

{\it WGA J0940$+$2603. $-$} The VLA A array map (Fig. 3f) shows two
extended lobes to each side of the core that partly overlap. The
western lobe contains an embedded broad jet-like structure, whereas
the eastern lobe appears to end in a hot spot.

\vspace*{0.1cm}

{\it TEX 0836$+$182. $-$} The VLA A array map (Fig. 8a) shows a
diffuse jet-like structure emerging from the core and stretching to
the south-east. Low-level extended emission on the scales of the jet
appears to be present everywhere around the core. The VLA C array map
(Fig. 8b) shows a well-collimated jet emerging from the core, which
extends to the south-west and therefore almost perpendicular to the
jet-like structure seen on smaller scales.

\vspace*{0.1cm}

In summary, the radio morphologies of all but two sources are
compatible with what is expected for a beamed FR II radio galaxy. In
particular, one-sided jets and extended halos surrounding the core-jet
structure are observed. Signs of a counterjet are detected in four
sources. The radio morphologies of WGA J1320$+$0140 and PMN J0630$-$24
are inconclusive. Since these sources are the only ones with elongated
cores and have the highest redshifts, radio maps of higher resolution
might reveal also in them one-sided jets. The observed broad jet-like
structures might then be rather lobes than large-scale diffuse jets.

\subsubsection{BL Lacs with FR I/II Extended Radio Powers}

In our sample, 25 sources have extended radio powers in the regime
where both FR I and FR II radio galaxies are found ($10^{24.5} <
L_{\rm ext} < 10^{26}$ W Hz$^{-1}$). About a third of these objects
(7/25 sources) show a radio morphology expected for a beamed FR II
radio galaxy. In the following we describe them individually and in
decreasing order of the extended radio luminosity.

\vspace*{0.1cm}

{\it WGA J1744$-$0517. $-$} The VLA A array map (Fig. 5a) shows a
one-sided jet that emerges from the core and stretches to the
north. The core-jet structure is embedded in lobe emission on scales
of a few arcseconds.

\vspace*{0.1cm}

{\it WGA J0528$-$5820. $-$} The ATCA 6A$+$6B array map (Fig. 2f) shows
two extended lobes to each side of the core. The southern lobe
contains an embedded broad jet-like structure, whereas the northern
lobe appears to end in a hot spot.

\vspace*{0.1cm}

{\it WGA 1202$+$44. $-$} The VLA A array map (Fig. 8d) shows the core
to be surrounded by a small ($\sim 10''$) diffuse halo. As the VLA C
array map (Fig. 8e) indicates, this halo is most likely the remainder
of an extended radio structure that is resolved out on these small
scales. The observed large-scale radio emission is indicative of an FR
II. Two lobes to each side of the nucleus are visible with the eastern
lobe containing a hot spot. The western lobe appears to have an
embedded one-sided jet, but a VLA B array map would be necessary to
confirm this.

\vspace*{0.1cm}

{\it B2 1147$+$245. $-$} The VLA A array map of \citet{Ant85b} shows a
one-sided jet-like structure emerging from the core and extending to
the south. A possible hot spot of a counterjet is seen on the opposite
side of the core.

\vspace*{0.1cm}

{\it WGA J1936$-$4719. $-$} The ATCA 6B array map (Fig. 6b) shows a
jet-like structure observed to continue over scales of tens of
arcseconds, possibly ending in a hot spot.

\vspace*{0.1cm}

{\it WGA J0245$+$1047. $-$} The VLA A array map (Fig. 1f) shows a
classical FR II structure, a one-sided jet emerges from the core in
direction south-east and a lobe with a hot spot is on each side of the
core. However, the VLA A$+$C array map (Fig. 2a) shows that this
source is rather an X-shaped radio galaxy
\citep[e.g.,][]{Cheung07}. Large-scale extended lobe emission is
visible to both sides of the core and almost orthogonal to the
small-scale lobes, thus giving the appearance of an X-shape.

\vspace*{0.1cm}

{\it WGA J1311$-$0521. $-$} The VLA C array map (Fig. 4d) shows
extended emission on both sides of the core. A lobe with an irregular
structure is present on the western side. Since the core appears to be
elongated in its direction, this lobe might contain a one-sided jet. A
VLA B array map would be necessary to confirm this. The structure on
the eastern side of the core is difficult to classify. It might be
either a highly irregular lobe with a hot spot or a strongly bended
large-scale jet.

\vspace*{0.1cm}

Of the remaining 18 sources in this luminosity range, five objects,
namely, WGA J0533$-$4632 (Fig. 3a), WGA J1539$-$0658 (Fig. 4f), WGA
J1834$-$5948 (Fig. 5d), WGA J1843$-$7430 (Fig. 6a), and WGA
J2330$-$3724 (Fig. 6e), four of which have been observed with the
ATCA, are barely resolved on the scales imaged and their association
with an FR class remains uncertain. The other 13 sources are most
likely beamed FR I radio galaxies for reasons as follows:

\noindent
$-$ four sources, namely, EXO 0556$-$3838 (Fig. 6f), PKS 2316$-$423
(Fig. 7c), RX J09168$+$523 (Figs. 7e and 7f), and 1ES 1212$+$078
\citep{Gir04}, show relatively broad one-sided structures, which seem
to be a mixture between a jet and a lobe. This structure is
reminiscent of the main characteristic of FR I radio galaxies, which
are jets that diffuse already at locations close to the core;

\noindent
$-$ seven sources, namely, WGA J0032$-$2849 (Figs. 1a and 1b), WGA
J0431$+$1731 (Fig. 2d), WGA J0449$-$4349 (Fig. 2e), WGA J1834$-$5856
(Fig. 5c), WGA J2258$-$5525 (Fig. 6c), EXO 1811$+$3143 (Fig. 7a) and
ON 325 \citep{Ant85b}, have a core surrounded by a smooth extended,
almost circular halo without the embedded one-sided jets typical of FR
IIs;

\noindent
$-$ and finally, the two sources with the lowest extended radio powers
and close to the FR I regime, namely, WGA J0428$-$3805 (Fig. 2c) and
WGA J0624$-$3229 (Fig. 3d), show two well separated radio lobes
oriented to the same side of the core and almost perpendicular to each
other. This radio morphology is reminiscent of a beamed wide-angle
tail (WAT) radio galaxy \citep[e.g.,][]{ODon90}, which is usually
grouped under FR I.

\subsection{An Inconsistency in the Classification Scheme} \label{incon}

We have shown that {\it (at least) a third of the DXRBS BL Lacs are
beamed FR II radio galaxies.} In the following we argue that an
heterogeneous parent population for BL Lacs is {\it expected} given
the current classification scheme for radio-loud AGN, which calls for
a revision of unified schemes.

Our present classification scheme has a general inconsistency, which
is illustrated in the diagram below. We separate radio galaxies and
blazars into their subclasses based on different criteria, namely,
radio morphology (and so radio power) and emission line strength,
respectively. However, observations of {\it both} beamed and unbeamed
sources suggest that these two criteria are not equivalent. Among the
radio galaxies, FR Is appear to form a homogeneous class with respect
to emission line strength, they have no or only weak emission lines
\citep[e.g.,][]{Owen95, Zir95}. On the other hand, FR II radio
galaxies appear to form a heterogeneous class. Whereas the large
majority have strong emission lines, a considerable number of FR IIs
exist that have weak emission lines \citep[e.g.,][]{Lai94, Zir95,
Tad98}. Then, since, by definition, BL Lacs are those blazars with
weak emission lines, their parent population is {\it expected} to be
heterogeneous. It will contain radio galaxies with no or only weak
emission lines, i.e., both FR Is and FR IIs. On the other hand, given
that only FR IIs have strong emission lines, the parent population of
radio quasars will be homogeneous.

\bigskip

\setlength{\unitlength}{0.75cm}
\begin{center}
\begin{picture}(12,4)
\thicklines
\put(0,1){\framebox(5,1){FR I radio galaxies}}
\put(5,1){\framebox(5,1){FR II radio galaxies}}
\put(2,0.3){\it separation by radio power}
\put(1.3,2.4){weak lines}
\put(5,2.4){weak lines + strong lines}
\put(0,3){\framebox(7.2,1){BL Lacs}}
\put(7.2,3){\framebox(2.8,1){quasars}}
\put(2.1,4.4){\it separation by emission line strength}
\end{picture}
\end{center}

\citet{L04} presented a classification scheme for radio-loud AGN that
correctly assigns the subclasses within the beamed and unbeamed
populations, and we have applied it to the DXRBS BL Lac sample (see
Section \ref{class} and Table \ref{general}, column (3)). Whereas for
9/44 sources their scheme is not applicable, the large majority (32/44
sources or 73 per cent) are classified as weak-lined and only three
sources as strong-lined radio-loud AGN. According to the
considerations above, the class of strong-lined radio-loud AGN should
comprise only FR IIs. Indeed, the three sources in this category,
namely, WGA J0245$+$1047, WGA J1840$+$5452 and 4C $+$55.17, have
extended radio powers and morphologies consistent with them being
beamed FR IIs. Of the nine sources without a classification, three
objects, namely, WGA J1320$+$0140, WGA 1012$+$06, and 3C 66A, have
radio properties compatible with FR IIs and could be strong-lined
radio-loud AGN. Candidates for this category are also the sources WGA
J0043$-$2638 and WGA J0100$-$3337, that have estimated extended radio
powers in the FR II regime, but are unresolved on the scales
imaged. The remaining four sources without a classification have radio
properties compatible with FR Is and are expected to be weak-lined
radio-loud AGN.

\section{Featureless BL Lacs}

In Section \ref{red} we have used the method of \citet{Pir07} to
determine lower limits on the redshifts of eight DXRBS BL Lacs. This
method takes into account that the optical spectrum of a BL Lac can
appear featureless if it is either at a low redshift ($z\la0.6$) and
strongly beamed or at a high redshift, in which case it can be also
moderately beamed. In this section we want to investigate if the
observed radio morphologies support this method.

For the majority of featureless DXRBS BL Lacs (5/8 objects) the method
of \citet{Pir07} suggests that they are strongly beamed, whereas three
sources, namely, WGA J0940$+$2603, WGA J1231$+$2848, and WGA
J1539$-$0658, could be either moderately or strongly
beamed. Therefore, we expect the first group to have on average larger
values of the radio core-dominance parameter $R$, a quantity that is
believed to be a suitable beaming indicator, than the second
group. Indeed, the first group appears to be on average a factor of
$\sim 2$ stronger core-dominated than the second group ($\log R =
0.83\pm0.35$ and $0.50\pm0.35$, respectively), but given the small
number statistics this result remains suggestive.

It is worth noting that among the first group one source (WGA
J0449$-$4349) is lobe- ($\log R=-0.33$) and not core-dominated as
expected by \citet{Pir07}. However, since this source is a high-energy
peaked BL Lac (see Section \ref{lblhbl}), it is possible that its jet
emission strongly dominates the host galaxy at optical frequencies,
thus rendering the optical spectrum featureless, although it is weak
relative to the extended emission at radio frequencies
\citep{Gio02}. We note also that among the second group one source
(WGA J1539$-$0658) is strongly core-dominated ($\log R=1.15$), which
means that its redshift might be considerably higher than the derived
lower limit.

\section{Low- and High-Energy Peaked BL Lacs} \label{lblhbl}


\begin{figure}
\centerline{
\includegraphics[scale=0.43]{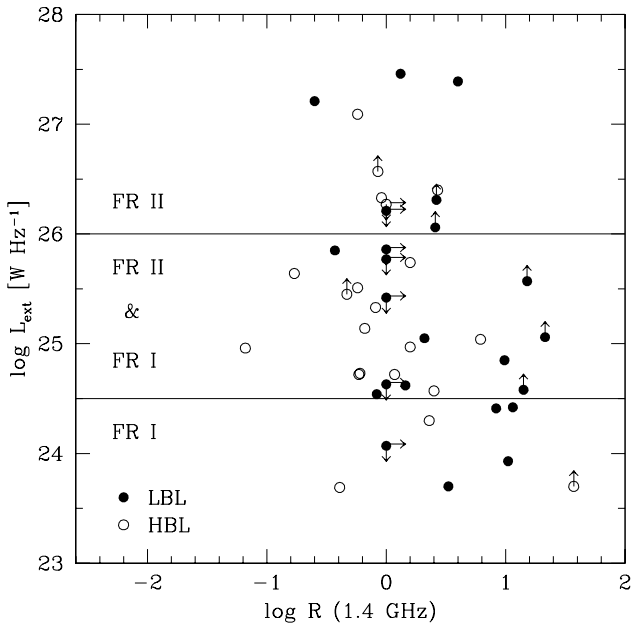}
}
\caption{\label{logRlextlblhbl} As in Fig. \ref{logRlext} for low-
(LBL; filled circles) and high-energy peaked BL Lacs (HBL; open
circles) from the DXRBS.}
\end{figure}

Depending on their spectral energy distributions (SEDs) BL Lacs are
divided into low- (LBL) and high-energy peaked BL Lacs (HBL), i.e.,
sources with a jet synchrotron emission peak at IR/optical and UV/soft
X-ray frequencies, respectively \citep{P95}. So far, two explanations
have been proposed for the observed range of SEDs. The oldest
hypothesis suggests that BL Lacs have jets with frequency-dependent
properties, in particular the higher the frequency, the more isotropic
the emission \citep{Mar86, Ghi89, Cel93a, Georg98}. Then, a BL Lac
appears as an LBL and HBL when viewed at smaller and larger angles,
respectively. A competing scenario proposes that BL Lac jets have a
wide range of magnetic fields and/or speeds \citep{Gio94, Gio95, P95,
Sam96}. The so-called 'blazar sequence' builds on this scenario and
advocates that the higher the (intrinsic) jet power and the strength
of the external, nuclear photon field, the stronger the
Compton-cooling of the jet particles, and the lower the synchrotron
emission peak frequency \citep{Fos98, Ghi98}.

The competing explanations for the observed BL Lac SEDs relied mainly
on observations of the 1-Jy and EMSS samples. However, given their
widely different selection bands (radio and X-ray frequencies,
respectively) and (radio) flux limits, these two surveys have most
likely presented a strongly biased view. Since the DXRBS contains a
sizeable number of both BL Lac subclasses we can now compare for the
first time their radio properties relatively free of selection
effects.

In Fig. \ref{logRlextlblhbl} we have plotted the radio core-dominance
parameter $R$ versus the extended radio luminosity at 1.4 GHz for the
DXRBS LBL (23 objects, filled circles) and HBL (21 objects, open
circles). LBL and HBL have been defined by $\log L_{\rm core}/L_{\rm
x} \ge 6$ and $<6$, respectively, where $L_{\rm x}$ is the total X-ray
luminosity at 1 keV \citep{P96}. Fig. \ref{logRlextlblhbl} presents in
principle a powerful test for both the postulations of the blazar
sequence and the two competing scenarios explaining the BL Lac
SEDs. The extended radio power is the most suitable measure of {\it
intrinsic} jet power for blazars, and according to the blazar sequence
we expect LBL to have on average higher values than
HBL. Fig. \ref{logRlextlblhbl} shows that this is not observed in the
DXRBS. Whereas LBL reach a factor of $\sim 2$ higher extended radio
powers than HBL, their average values are very similar ($\log L_{\rm
ext}=25.35\pm0.23$ and $25.28\pm0.20$ for LBL and HBL,
respectively). We note that the quoted averages have been calculated
treating both lower and upper limits as detections, but a similar
result is obtained, if the limits are omitted. Similar to our
approach, \citet{Nie08} have recently pointed out that the blazar
sequence might be an artifact of relativistic beaming effects, showing
that it is not observed once the powers are corrected for Doppler
boosting. However, these authors find for BL Lacs an inverted blazar
sequence, which we cannot confirm.

The radio core-dominance parameter $R$ is believed to be a suitable
indicator of orientation (the higher its value, the smaller the jet
inclination angle), and, if orientation determines the SED differences
between LBL and HBL, we expect their average values to be
different. Fig. \ref{logRlextlblhbl} shows that LBL have on average a
factor of $\sim 3$ higher radio core-dominance parameters than HBL
($\log R=0.53\pm0.14$ and $0.00\pm0.13$, respectively, calculated
omitting the limits). This result is similar to what has been found
previously for comparisons between LBL from the 1-Jy survey and HBL
from the EMSS \citep[e.g.,][]{Rec01} and is usually interpreted in
support of the 'different orientation' theory for the BL Lac SEDs.

Nevertheless, selection and definition effects can account for this
result. Since, by definition, LBL have on average higher radio {\it
core} luminosities (and lower X-ray luminosities) than HBL and in the
DXRBS both span a roughly similar range in {\it extended} radio power
(see above), LBL are bound to have higher $R$ values than HBL. In
other words, in a radio-flux limited sample, such as, e.g., the DXRBS,
LBL and HBL {\it are defined} to be those sources with on average
higher and lower radio core-dominance parameters, respectively. In
order to determine if the relative orientations of the two BL Lac
subclasses are indeed different, one needs to use instead an
orientation indicator, whose definition is independent of the LBL-HBL
one. Such a study was done by \citet{L02}, who used the Ca H\&K break
value (a stellar absorption feature in the optical spectrum) as an
orientation indicator, and these authors found no significant
differences between LBL and HBL, in support of the claims of
\citet{P95}.

\section{Summary and Conclusions}

Our knowledge of the radio properties of BL Lacs is based mainly on
the 1-Jy and EMSS samples. However, these surveys have presented a
biased view of BL Lac physics. Therefore, we have obtained deep radio
images of a complete sample of 44 BL Lacs selected from the Deep X-ray
Radio Blazar Survey (DXRBS). We have observed the northern sources
with the VLA in both its A and C configurations and the southern
sources with the ATCA in its largest configuration. Our main results
can be summarized as follows.

(i) Current unified schemes identify the parent population of BL Lacs
with FR I radio galaxies, however, in recent years evidence has
accumulated that some BL Lacs might in fact be relativistically beamed
FR II radio galaxies. We find that also in the DXRBS, based on both
the extended radio powers as well as the radio morphologies, (at
least) a third of the BL Lac sample can be identified with
relativistically beamed FR IIs.

(ii) We discuss an inconsistency in the current classification scheme
for radio-loud AGN, which explains why FR II-BL Lacs are in fact {\it
expected} to exist. This inconsistency emerges since we separate radio
galaxies and blazars into their subclasses based on different
criteria, namely, radio morphology (and so radio power) and emission
line strength, respectively. However, these two criteria are not
equivalent.

(iii) The so-called 'blazar sequence' posits that the amount of
Compton cooling of the jet particles determines the frequency of the
synchrotron emission peak \citep{Fos98, Ghi98}. In particular, it
expects low-energy peaked BL Lacs (LBL) to have on average higher {\it
intrinsic} jet powers than high-energy peaked BL Lacs (HBL). We
compare the extended radio powers of the DXRBS LBL (23 objects) and
HBL (21 objects) and find that, contrary to the expectations of the
blazar sequence, their average values are similar ($\log L_{\rm
ext}=25.35\pm0.23$ and $25.28\pm0.20$, respectively).

We are currently extending our radio observation program for the DXRBS
BL Lac sample to investigate smaller-scale source structure. In a
future paper we will present results from VLBI.

\section*{Acknowledgments}

H.L. thanks the astrophysics department of Oxford University for its
hospitality during the last months of this work. The VLA is operated
by the National Radio Astronomy Observatory, which is a facility of
the National Science Foundation operated under cooperative agreement
by Associated Universities, Inc. The ATCA is part of the Australia
Telescope, which is funded by the Commonwealth of Australia for
operation as a National Facility managed by CSIRO.

\bibliography{references}

\bsp
\label{lastpage}

\end{document}